\documentclass[aps,floatfix,preprint]{revtex4}

\usepackage{amsmath}
\usepackage{graphicx}
\usepackage{color}
\usepackage{pslatex}
\usepackage{setspace}
\usepackage[normalem]{ulem}

\begin{document}

\title{Quantum Thermodynamics, Entropy of the Universe, Free Energy, and the Second Law}

\author{George L. Barnes}
\affiliation{Department of Chemistry and Biochemistry,\\Siena College\\Loudonville, NY 12211, USA}
\author{Phillip C. Lotshaw}
\author{Michael E. Kellman}
\affiliation{Department of Chemistry and Biochemistry and Institute of Theoretical Science, University of Oregon \\ Eugene, OR 97403, USA}
\date{\today}

\begin{abstract}

We take the view that the standard von Neumann definition, in which the entropy $S^{vN}$ of a pure system-environment state $| \Psi_{SE} \rangle $  is zero, is in evident conflict with the  statement of the second law that the ``entropy of the universe" $S_{univ}$ increases in spontaneous processes, $ \Delta S_{univ} > 0$.   Here we seek an alternative entropy of the universe  $S_{univ}^Q$ that is in accord with the second law, in a spirit not dissimilar to von Neumann himself in lesser-known work.  We perform simulations of time dependent dynamics for a previously developed \cite{polyadbath} model  quantum system becoming entangled with a quantum environment.  We test the new definition of the entropy of the system-environment ``universe" against the standard thermodynamic relation $  \Delta F_{sys} = - T \Delta S_{univ}  $, calculating system properties using the reduced density matrix and standard von Neumann entropy.   Generally good agreement is obtained, showing the compatibility of an entropy for a pure state of a universe with the  statement of the second law and the concept of free energy.    Interesting deviation from microcanonical behavior within the zero order energy shell is observed in  a context of effectively microcanonical behavior within the much larger total basis of the time dependent universe.

\end{abstract}

\maketitle

\section{introduction}

In recent years, a  number of authors \cite{tasaki1998, Gemmerarticle, Gemmer:2009, Popescu2006, Popescu2009, Goldstein2006, vNcommentary, Goldstein2010, Goldstein2015, vNtrans, Reimann2008, Reimann2016, Rigol, Belgians2, Belgians, PolkovnikovicEntropy,  HanEntropy,   polyadbath, Reeb} have launched a re-examination of the foundations of quantum thermodynamics.  One prominent line of thought has explored the idea that the approach to thermodynamic equilibrium has its physical origin in entanglement of a quantum system and environment.  The system and environment SE are often taken to be an isolated ``universe"  in a pure quantum state  $ |\Psi_{SE} \rangle$,  starting in a separable SE state   and evolving in time with entanglement toward thermal equilibrium.  Furthermore, the thermodynamics of truly large and even cosmic scale pure state entities is of great interest, e.g. some treatments of the quantum thermodynamics of black holes including the ``black hole firewall paradox" \cite{AMPS,Bousso},  and even of the entire universe in very different models \cite{WaveFunctionUniverse,steinhardtturok,Rovelli}.     The idea of the quantum thermodynamic evolution of a system environment universe state   $ |\Psi_{SE} \rangle$, of whatever scale, stimulates the investigation here of what seems  a  natural question, leading us to introduce the idea (to be given a precise definition later)  of a total or universe quantum entropy $S^Q_{univ}$ for an SE pure state, as follows.  In standard thermodynamics, the second law is expressed as the statement that, except  at equilibrium or during transient  fluctuations, the entropy of the universe is always increasing: 

\begin{equation} \Delta S_{univ} > 0 \label{secondlaw}     \end{equation}

\

\noindent Further, for a process with zero pressure-volume work, 
a very important statement in thermodynamics relates the free energy change of the {\it system} at fixed $T, V$ to the entropy change of the {\it universe}:

\begin{equation}   - \frac{1}{T}    \   \Delta F_{sys} =  \Delta S_{univ}          \label{greatresult}  \end{equation}

\noindent The free energy change of the system is thus a surrogate for the entropy change of the universe in the  second law.   This is the meaning of the statement that the free energy $F$ is a thermodynamic potential at fixed $T$ and $V$.

These considerations raise a problem in quantum thermodynamics.   In the most common definition, introduced by von Neumann in his famous book of 1932 (the English translation  was reissued \cite{vNbook} in 1996), the quantum entropy is given in terms of the eigenvalues $\rho_i$ of the density matrix:

\begin{equation}  S^{vN}  = - \sum \rho_i  \ln \rho_i    \label{vNe}  \end{equation}

\noindent where the $\rho_i$ are probabilities summing to unity.  By this definition, the entropy of a pure state is zero.  Then   if the combined SE universe is described by a density matrix $ \rho_{SE} $  for a pure state, we have $S^{vN}_{univ} = 0$.    Hence, the von Neumann definition of the quantum entropy does not encompass the idea of the entropy of a pure state SE closed universe, apart from quantum entanglement with some further exterior world.  (In this paper, we use $sys$ and $univ$ to refer generically to system and universe; we use S, E, and SE when referring to the specific model system S, environment E, and  universe SE.) 
The statement  of the second law in standard thermodynamics,  given in Eq. \ref{secondlaw}, thus ceases to have meaning, along with the foundation in  the second law of the standard relation   between free energy and spontaneous processes Eq. \ref{greatresult}, and the basis of the theory of equilibrium in the law of van't Hoff.    This is not just a matter of settled results in thermodynamics: in modern treatments of nonequilibrium thermodynamics e.g. Kondepudi and Prigogine \cite{kp}, the entropy change of the universe regarded as ``irreversible entropy production" is taken as a basic working idea.  At the cosmic scale, the presumed lack of a meaningful entropy for a pure state would seem to be a gap in the description e.g. of black hole thermodynamics, to say nothing of the entropy of the universe as a whole.

With this in mind we ask, can we define a notion of the entropy of the universe $S^Q_{univ}$ that does not have the problem of the von Neumann definition?  And will such a definition give results in accord with Eq. \ref{greatresult} in numerical simulations of quantum systems?  Our point of departure is a recent paper \cite{polyadbath}, following work of Gemmer et al. \cite{Gemmerarticle,Gemmer:2009} that  simulated a small quantum system embedded in a quantum environment with a view toward comparison with thermodynamic behavior.   The time dependent dynamics of the system were revealed through the calculation of the reduced density matrix and subsequent calculation of the von Neumann entropy, system populations, and spatial density plots.  It was found that various initial states of a coupled oscillator system approach the same final distribution, which can be characterized as fluctuation about a Boltzmann distribution with a temperature $T$.  This is fully in accord with recent analytical work \cite{tasaki1998, Reimann2008, Popescu2009, Goldstein2010, Goldstein2015, vNcommentary, vNtrans,Reimann2016} on the evolution of an SE universe toward ``typical" states that mirror the microcanonical and canonical ensemble description of a system embedded in an environment.  Our focus here is on the question of the entropy of the total SE universe in this time evolution.   As will be developed in detail, we define a quantum entropy for a pure SE state as the standard Shannon entropy defined with respect to the zero-order energy basis $\{| \alpha \rangle \} = \{ | s \rangle |\epsilon \rangle \}$  of the SE complex:

\begin{equation} | \Psi_{SE} (t)  \rangle = \sum_{\alpha } c_{\alpha }  (t) | \alpha \rangle.    \label{expansion}  \end{equation}  

\noindent   Taking

\begin{equation}  p_{\alpha} (t) = | c_{\alpha} (t) |^2    \label{probability}    \end{equation}

\noindent we define the entropy of the universe

\begin{equation}  S^Q_{univ}  = - \sum_\alpha p_{\alpha} \ln p_{\alpha}     \label{sunivSE}  \end{equation}    

\noindent Our strategy is to calculate $ - T \Delta S^Q_{univ}$, and compare this with a separate calculation of $ \Delta F_{sys}$.      To the extent that these two quantities are equal, we will obtain a recovery of standard thermodynamics ideas and results.  The general method gives  $\Delta S_{sys}$ for various initial states as the change in the von Neumann entropy  $\Delta S^{vN}_{S}$ calculated from the reduced density matrix of the system, and a suitably defined temperature $T$.  Accordingly, for the thermalization process, we  calculate the free energy change  $\Delta F_{sys} = \Delta E_{S} - T \Delta S^{vN}_{S}$.

It is important to be clear what we mean with the terms ``system, environment, universe."  We have in mind primarily a particular small system S, of laboratory dimensions or less; and a corresponding environment E.  Together, the bipartite quantum system-environment SE makes up the ``universe."  This certainly is not necessarily meant to imply cosmic dimensions.  In fact, we see the most likely direct interest of the work here in its possible significance for very small quantum ``universes" that might exist in the laboratory or in nature, isolated from their larger surroundings on a sufficiently limited timescale.    However, many of the same questions that motivate us might apply to systems of cosmic size.  It is conceivable that some of our considerations could shed light on thermodynamic problems in cosmological systems, though we will not pursue such possibilities here.

  It could be argued that a focus on the ``universe" entropy is misplaced, because any particular system-environment SE will be embedded within a larger SEE$'$, with the smaller universe SE becoming rapidly entangled with E$'$, so that SE can be treated by means of its von Neumann entropy, and so on with ever-expanding layers of entanglement with the environment.  Such an approach with a mixed state SE has been developed by Esposito et al. \cite{Belgians} and by Reeb and Wolf \cite{Reeb}.  On the other hand, this necessity of entanglement for $S^{vN}$ can be seen as simply begging the question of the entropy of the pure state of the ``universe," however narrowly or widely defined, in the ``individualist" approach \cite{vNcommentary} that emphasizes a pure state of an SE total system.  Furthermore, it is perfectly reasonable to consider the question of timescales.  Small universe situations can be imagined and perhaps constructed and probed in laboratory experiments, in which SE  thermalizes  far more rapidly than entanglement occurs with the next environment layer in SEE$'$.  A pure state entropy $S_{univ}^Q$ could be very useful in these situations.

   Some readers may be uneasy about defining a new entropy $S^Q_{univ}$ without   justification in prior established principles.  It may be good to keep in mind  Einstein's injunction, stated in its most widely quoted instance in the popular book with Infeld \cite{EinsteinInfeld}:  ``Physical concepts are free creations of the human mind, and are not, however it may seem, uniquely determined by the external world."  This seems particularly apt when considering entropy, which to an unusual degree might appear to have the character of a contrivance that is nonetheless accepted because of its usefulness in describing observed behavior of the natural world.  This can pertain to the classical entropy of thermodynamics, the statistical entropy of Boltzmann, and the von Neumann quantum entropy as well as the universe quantum entropy $S^Q_{univ}$ defined and investigated here.  The ``validity" of the latter concept, like its predecessors, will depend on its usefulness in accounting for natural phenomena.

\section{Historical Background and Contemporary Considerations. }

In this section we present considerations on  connections of the entropy $S^Q_{univ}$  of this paper to contemporary work of a number of authors, but also going back to some lesser-known work of von Neumann.     
As noted already, a number of authors \cite{tasaki1998,Gemmerarticle, Rigol, Reimann2008, Reimann2016, Popescu2009, Goldstein2010, Goldstein2015, vNcommentary, vNtrans} have investigated the time evolution of a state    $| \Psi_{SE} \rangle $   for a system-environment  complex, often explicitly designated  a ``universe" but which might correspond to a small isolated ``total system," just as here.  The goal is to show that the time evolution {\it of an individual quantum pure state}  leads naturally to thermodynamic behavior, i.e. thermalization of the system S to a standard Boltzmann distribution.  Most of this work has not considered an entropy associated with the pure state $| \Psi_{SE} \rangle $, e.g. Ref. \cite{Popescu2006} emphasizes that the von Neumann entropy of the pure state universe is zero.  However, there have been some exceptions -- including von Neumann himself!  

In an old paper    \cite{vN}   that has become better known with its recent translation into English \cite{vNtrans}   along with a lucid modern commentary \cite{vNcommentary}, von Neumann sought to deal with the problem of extending quantum statistical mechanics, including the notion of an {\it entropy of a pure state}, to recover results of the standard microcanonical ensemble approach in statistical thermodynamics.  He was concerned as here with  explaining thermodynamic behavior in terms of the time evolution of a pure quantum state for a closed system.  This line of thought  does not however appear in von Neumann's later book \cite{vNbook} on the mathematical foundations of quantum mechanics.  But more recently,  parts of this  program have been taken up again in   contemporary fundamental investigations of the foundations of quantum statistical mechanics.

In  Ref. \cite{vN,vNtrans} von Neumann recognizes a tension between what the commentary Ref. \cite{vNcommentary} characterizes as   ``individualist" and ``ensemblist" points of view in quantum statistical mechanics.   In classical statistical mechanics, the individualist point of view corresponds to the idea of a single detailed complex system time-evolving according to deterministic dynamics to give thermodynamic behavior, including an idea of an entropy.  In contrast, the ensemblist point of view encompasses the use of both the microcanonical and canonical ensembles to describe thermodynamic behavior.   In classical statistical mechanics the individualist and ensemblist views can be reconciled with more or less conviction, as was in effect argued by Boltzmann, Gibbs \cite{Gibbs1902} and Einstein \cite{Einstein1902,Einstein1903,Einstein1904,Damour:2006}.  Ref. \cite{vNcommentary} asserts that a goal of von Neumann \cite{vN,vNtrans} is to reconcile the two points of view in quantum statistical mechanics.

 The quantum individualist point of view starts from a single pure state, corresponding roughly to the idea of a single classical trajectory in phase space.  von Neumann was concerned in Ref. \cite{vN,vNtrans}  to show that an individual pure quantum state would show time evolution compatible with  microcanonical behavior, and that this would be consistent with canonical behavior of the embedded system. Of interest to us here, in Ref. \cite{vN,vNtrans}, von Neumann also claimed to demonstrate a ``quantum H-theorem" related to an entropy for a pure quantum state.  This entropy is completely different from the von Neumann entropy for a mixed state, as von Neumann is careful to emphasize.    It should be noted  emphatically that the entropy $S^Q_{univ}$ for a pure state that we will propose here  is different conceptually and computationally from von Neumann's entropy for a pure state.  His pure state entropy is based on an idea of coarse graining the state probabilities into distinct ``macroscopic" subspaces and then using the course grained probabilities to calculate an entropy.  His method trivially gives the Boltzmann entropy $S = k \ln W$ when the subspaces are occupied with microcanonical probability, independent of the details of the pure state.  On the other hand, our entropy will be defined without any coarse graining and we will have occasion  to investigate significant deviations of our entropy from the Boltzmann entropy relation.

This distinction aside, the investigation here of a new system-environment entropy $S^Q_{univ}$ can be regarded  as fitting broadly with the goals of von Neumann's work as well as contemporary endeavors.   Several of these have focussed on new ideas about entropy, related in various ways to what we are trying to do.  Han and Wu \cite{HanEntropy} introduced a pure state entropy defined with respect to a ``quantum phase space" basis, and argued that  under certain conditions this entropy will typically increase to its equilibrium value.  Kak \cite{KakEntropy} has discussed the information content of pure quantum states in the context of communication between a source and receiver, with an information content related to an entropy for a pure state.  Reimann  \cite{Reimann2016} and Goldstein et al. \cite{Goldstein2015} considered a thermodynamic entropy based on a count of states.     Polkovnikov \cite{PolkovnikovicEntropy} defined a ``diagonal-entropy" using the energy eigenstate probabilities of a closed system and studied its thermodynamic properties. In this formalism entropy changes due to changes in the Hamiltonian e.g. after the removal of a constraint, however this precludes the possibility of entropy production in isolated pure state systems which have time-independent Hamiltonians.  More similar in aim to our work, but not so much in approach, Esposito et al. \cite{Belgians}  and also Reeb and Wolf \cite{Reeb} have defined an ``irreversible entropy production"   $\Delta_i S$  (following the terminology of Prigogine \cite{kp}, essentially the ``entropy change of the universe")  related to the change in system von Neumann entropy $\Delta S^{vN}$ as it exchanges heat $Q$ with an environment that begins in a canonical mixed state.  With respect to the latter aspect, their approach does not lead to an entropy change of an SE universe in a pure state, as we seek here.  To our knowledge, none of these other works, including that of von Neumann, have  proposed the specific entropy of Eq. (\ref{sunivSE}) that we investigate here.    The introduction here of a quantum $S^Q_{univ}$ is intended as a distinctive approach that could lead in directions that transcend classical statistical mechanics.

\section{spontaneous process and entropy in a quantum universe}

In classical statistical mechanics (including early quantum statistical mechanics of quantized energy levels, but not quantum states undergoing entanglement) it is typically assumed that spontaneous processes occur in an SE universe when a constraint is removed.  This allows a process that might be termed ``ergodization" (however loosely defined) but might better be called ``microcanonical spreading" (to avoid any connotation of time-averaging; see comments shortly below)  in which the universe attains the maximal entropy $S = - \sum p_\alpha \ln p_\alpha = \ln W$ according to the Boltzmann microcanonical definition $p_\alpha = 1/W$, where $W$ is the number of states in the microcanonical ensemble.  Here we want to explore thermodynamic behavior of a spontaneous process in a system-environment quantum universe.  We will focus on perhaps the most basic spontaneous process:  heat flow between a quantum system and a quantum environment, with no mechanical work.  In this section, we propose a definition for a quantum entropy for the SE universe, discuss the rationale for this definition, and finally specialize to the particular process for which we will perform quantum simulations.

\subsection{Reference Basis for Heat Flow and Quantum Entropy of the Universe }      \label{referencebasis}

We now propose the definition of the quantum  entropy of the universe $S^Q_{univ}$.   We think of a universe described by a pure state (this last condition could be generalized, but that is not our purpose here),  in general entangled, of a system S and an environment E.  (Everything goes through when the universe consists of just a simple system S.)  To define an entropy  $S_{SE} = S^Q_{univ}$  we choose a ``reference basis"  $\{ | \alpha \rangle \}$.  In this basis a pure state is expressed as

\begin{equation} | \Psi_{SE} (t)  \rangle = \sum_\alpha c_\alpha (t) | \alpha \rangle.    \label{expansionagain}  \end{equation}  

\noindent   Then taking

\begin{equation}  p_\alpha (t) = | c_\alpha (t) |^2    \label{probability}    \end{equation}

\noindent we define the entropy of the universe

\begin{equation}  S^Q_{univ} = S^{\{\alpha\}} _{univ} = - \sum_\alpha p_\alpha \ln p_\alpha     \label{suniv}  \end{equation}

\noindent \textit{with respect to the reference basis} $ \{ | \alpha \rangle \}$. 
Note that the quantities $p_\alpha$ come from the coefficients of the pure state, not from the eigenvalues of the universe density matrix $ \rho_{SE}$, which of course would yield zero in Eq. \ref{suniv}.  
As already specified with Eqs.  (\ref{expansion}-\ref{sunivSE}), we choose as reference basis the  zero-order energy basis $\{| \alpha \rangle \} = \{| s \rangle |\epsilon \rangle \}$  of the SE complex.

 What is the meaning of this procedure?  The expression for the entropy on the right-hand side of  Eq. (\ref {suniv}) is not new. It has an evident relation to the Shannon information entropy.   In the quantum context it has been discussed as the ``conditional information entropy" by Stotland et al.  \cite{informationentropystotland}.   
This entropy depends on the choice of reference basis.  In fact, it is trivial to pick a ``bad" reference basis, such that the entropy $S^Q_{univ}$ never changes.  The basis of energy eigenstates of the SE universe has this property, because any time-dependent state in this basis has constant coefficients $c_\alpha$ in (\ref{expansion}), hence constant entropy.   In fact, most possible reference basis sets are similarly ``bad," with $\Delta S^Q_{univ} \approx 0$.

We need some criterion for picking a ``good" reference basis for defining a meaningful thermodynamic entropy and its change.  We posit that the reference basis for $S^Q_{univ}$ in the process of thermalization should be identifiable or compatible with a (possibly macroscopic) property or set of properties that would be observed for the system S and possibly the environment E.    Here we will be simulating a quantum total system SE in which energy flows between a system S and an environment E -- a process of heat flow.  We take the reference basis to consist of microscopic ``cells" of the relevant ``macroscopic" states of the system, here, simply the energy.  Hence, we will take the cells to be basis states of the tensor product space of the zero order energy eigenstates of S and E.

We can understand the rationale for this further, as follows.     We typically would measure an S energy level, e.g. the energy  of a Brownian particle in a gravitational field.  Then, if we are concerned with thermalizing energy flow,  the most natural further observation would be of the zero-order energy of E to give a total zero order energy of SE.  Repeated on an ensemble, this would give  our $S^Q_{univ}$ as an actual von Neumann entropy of SE, called the ``mixing entropy" \cite{Wehrl},  the result of a complete measurement on an ensemble in a certain basis, i.e. here the zero order SE basis.  We do not foresee actually measuring the E zero order energy, though that is not inconceivable in an experiment on a sufficiently small SE ``universe." Rather, we think of the E zero order basis as most natural in the context of thinking about thermal processes where the S energy would be measured and would correlate with E.  In fact, if one tried to measure a property of E corresponding to a basis that was drastically different from the E zero order energy basis, e.g. a basis of superposition states of very different E zero order energies,  one would cause a gross disturbance to the SE total system by destroying the correlation between S and E energies that is normally present in thermalization. That is, choosing the E zero order energy basis or something very like it  would seem to cause the least disturbance to the SE thermal state  if the mixing entropy were actually measured.

\subsection{Thermalization, Microcanonical Ensemble, and  {Quantum State Fragmentation}  }      \label{tmef}

What should we expect of the quantum thermodynamics in this reference basis?  We will argue by analogy to the basic ideas of classical statistical mechanics.   There we expect the SE universe to be described by the microcanonical ensemble.   We expect the system S embedded in E to reach a thermal distribution.  This takes place through a process of  {microcanonical spreading or fragmentation} in which the microstates of the microcanonical ensemble become, at least for practical purposes, equally likely (in the chosen reference basis, here the zero-order energy states).

Let us consider how these notions carry over to a fully quantum mechanical universe SE of a system S becoming entangled with the environment E.  First consider the question of thermalization.  The S and E energies are properties of S and E considered separately that one might observe or measure. When we treat  entanglement of S with E in the process of heat flow, then if thermal conditions prevail,   a Boltzmann distribution of  zero order energy states of S should emerge in the reduced density matrix for the system.  Then the zero-order energy basis of the system diagonalizes the reduced density matrix, i.e. the S energy basis is carried over to the Schmidt SE basis \cite{nielsenchuang}.   In fact, in Ref. \cite{polyadbath}, this is exactly the kind of behavior, subject to fluctuations, found for the system and environment that we use in the present paper.    This is an indication that the system energy basis is a ``good" basis for the system von Neumann entropy, and also good  as a component of the reference basis for the universe entropy.

The issues of how thermalization relates to microcanonical behavior in a quantum context, let alone how this might relate to a notion of  {``ergodic"} behavior, are more subtle and problematic.  Classically,  microcanonical behavior is constituted by having equal probabilities $p_\alpha = 1/W$ within a microcanonical energy shell of $W$ states of equal energy.  This already assumes a division into quantum states, but not quantum dynamics.  The assumption of equal probabilities is often justified by appeal to an ergodic-type hypothesis e.g. that time average is effectively equal to the ensemble average.  This is problematic in our approach in which the $p_\alpha$ values are taken as instantaneous values in Eq.  \ref{probability}, $p_\alpha = |c_\alpha (t)|^2$.  There is no time average here, and certainly no a priori justification for assuming that all the instantaneous $p_\alpha$ values are equal.   Furthermore,  ``microcanonical" behavior is an ambiguous concept in quantum thermodynamics.  Conventionally, a microcanonical ensemble depends on having a narrow energy shell.  However, in quantum mechanics, a time-dependent state in a spontaneous process does not have a well-defined energy.  The $W$ microcanonical states that are supposed to have equal probabilities are thus also a problematic idea.  Despite these necessary conceptual distinctions,  we will proceed with the definition of $S^Q_{univ}$ of Eq.  \ref{suniv}, examining  whether Eq. \ref{greatresult}  holds in empirical simulations, and the extent to which   microcanonical fragmentation behavior is observed.

\section{Computational Quantum Simulations} 

Our quantum simulations model a system S and environment E which become entangled in the course of the dynamics of the universe SE.   Here we outline the features of the model.  More details are presented in Ref. \cite{polyadbath} where we found that the model universe evolves toward a thermal distribution with a temperature $T$.  This is consistent with computational studies of thermodynamics in a variety of other models \cite{Gemmerarticle, Olshanni, Rigol, Belgians2}.

 Our system consists of two linearly coupled harmonic oscillators; the environment E consists of a bath of levels with a degeneracy pattern devised to have a statistical mechanical temperature.  We choose a random coupling scheme between S and E.  For the system S, we calculate the reduced density matrix from the time-evolving pure state of the SE universe.    From these we obtain the free energy change $\Delta F_{sys}$ of the system and the entropy change $\Delta S^Q_{univ}$ of the universe.

\subsection{System, Environment, Model Hamiltonian, and Temperature \label{sec:method:model}}

The total Hamiltonian operator is a sum of three parts

\begin{equation}
\hat{\mathrm{H}} = \hat{\mathrm{H}}_{S} + \hat{\mathrm{H}}_{E} + \hat{\mathrm{H}}_{SE}  \label{SEmodel}
\end{equation}

\noindent for system, environment, and system-environment interaction.    We work in the basis of the energy eigenrepresentation of both the system and environment which means that both $\hat{\mathrm{H}}_{S}$ and $\hat{\mathrm{H}}_{E}$  are represented in diagonal form.  
The system basis will consist of states  $\{ \vert s \rangle \} = $ $\{ | n \rangle \}$; the environment basis  of states $\{ \vert \epsilon \rangle \} = $ $\{ | m, l  \rangle \}$ with quantum numbers $n, m, l$ to be defined shortly.  The product basis is then  $ \{ |n \rangle \otimes |m,l \rangle \equiv |n,m,l \rangle \}$.

\subsubsection{Isolated System Hamiltonian \label{isolatedsystem}}

For the system we take two linearly coupled oscillators, labeled 1 and 2, with Hamiltonian

\begin{equation}   \hat{\mathrm{H}}_{S} = (n_1 + n_2) \omega_0 + \kappa (a_1^{\dagger} a_2 + a_1 a_2^{\dagger})  \label{system}  \end{equation}  

\noindent where  $n_1, n_2$ are the numbers of quanta in  modes 1, 2 and $\omega_0, \kappa$ are parameters that we take to be 34.64 and 1.0 in reduced units (the rationale for various parameter choices is detailed below ).   The coupled system yields normal mode eigenstates that can be labeled by quantum numbers $n_s, n_a$ for the number of quanta in the symmetric and antisymmetric  modes.  The Hamiltonian in this normal mode representation is

\begin{equation}  \hat {\mathrm{H}}_S = n_s \omega_s + n_a \omega_a = n_s  (\omega_0 - \frac{1}{2} \kappa) + n_a (\omega_0 + \frac{1}{2} \kappa)     \end{equation}  

\noindent The coupling in Eq. \ref{system} preserves the total quantum number $N = n_1 + n_2 = n_s + n_a$, often referred to as the polyad number.  Associated with a given $N$ are a set of $N+1$ normal mode states, referred to as a polyad of states.  Each distinct polyad constitutes an isolated system since the Hamiltonian preserves the polyad number (i.e. distinct polyads are not coupled).  The normal mode energies are equally spaced within a polyad.    In reduced units we define the value of the spacing between states within the $N = 5$ polyad as 1.  Although the calculations are performed in reduced units,  we present final results in wavenumbers and picoseconds based on the absolute spacing of 111.77 cm$^{-1}$ between these polyad states.  This value  corresponds to the parameter $\kappa$ in Eq. \ref{system} in absolute units  and was adapted from a fit to the water stretching mode spectrum in Ref. \cite{Xiao89sphere,Xiao90cat}.    For convenience we will label each energy eigenvalue of the normal mode system using the quantum number $n = 0 \cdots 5$.

\subsubsection{Isolated Environment Hamiltonian}     \label{isolatedenvironment}
We model the environment by making a rather drastic, but computationally straightforward approximation to a true continuum of environment levels, following our previous work \cite{polyadbath} and the work of Borowski et al. \cite{Gemmerarticle} and \cite{Belgians2} for alternate finite bath models. Specifically, we use a set of evenly spaced harmonic levels that can be completely defined using an energy quantum number $m$ and a degeneracy quantum number $l$.  These quantum numbers together specify each zero order E state  as $| \epsilon \rangle$ =  $|m,l\rangle$.    The zero order E Hamiltonian is taken as  

\begin{equation}
\hat{\mathrm{H}}_{E}^0 |m,l\rangle = m \omega_{E} |m,l\rangle
\end{equation}

\noindent where $\omega_{E}$ is a parameter of the model, taken to be 1.  We choose to make the zero order harmonic spacing of E equal to that of the system.  This is based on the idea that SE interactions that approximately conserve total energy will be the most favorable and important interactions that a system would have with a true continuum environment.  Ignoring the levels in-between allows us to make the computations tractable while maintaining the essential physics involved in SE energy transfer. The degeneracy of each level is given by $g(m)=Ab^{m\omega_E}$ where $A$ and $b$ are  parameters of the model. The degeneracy pattern is based on how the degeneracy of a true continuum environment increases with energy at a given temperature $T,$ as will be explained shortly.  In sum, the model environment is a computationally tractable approximation that is designed to mimic the most important thermodynamic properties of a true constant temperature environment heat bath: statistical degeneracies corresponding to a temperature $T$ and the ability to exchange heat with the system while conserving total energy.

\subsubsection{Temperature}

The thermodynamic definition of temperature is given by

\begin{equation}
\frac{1}{T} = \frac{\partial S}{\partial U}.
\end{equation}

\noindent The connection with statistical mechanics is made through the relation $S = k_b \ln W$, and then related to our construction by using the degeneracy formula  $S = k_b \ln\left ( Ab^U\right)$ where $k_b$ is the Boltzmann factor.  This yields  

\begin{equation}
T = \frac{1}{k_b\ln\left( b \right )}\label{eq:Temp}
\end{equation}

\noindent which shows that the base of the exponential scaling of the degeneracy in fact defines the temperature.  In this work we take $b = 2$ which leads to a temperature of 230.41 degrees Kelvin.  We assume that we work in a sufficiently narrow energy range of the bath that $T$ is energy independent.

\subsubsection{System-Environment Interaction}

The SE interaction will be taken to be a random coupling.    The SE interaction disrupts the perfect degeneracy pattern defined above.  We include the diagonal portion of the SE interaction directly in the E Hamiltonian and represent it as

\begin{eqnarray}
\hat{\mathrm{H}}_{E}^{shift}|m,l\rangle &=& X(m,l) |m,l\rangle\\
\hat{\mathrm{H}}_{E} &=& \hat{\mathrm{H}}_{E}^0 + \hat{\mathrm{H}}_{E}^{shift}  \label{H_E}
\end{eqnarray}

\noindent where $X(m,l)$ is a random variate selected from a gaussian distribution with zero mean and standard deviation,      $\sigma=\gamma\omega_{E}\sqrt{2}$ .  The eigenvalue equation for the final E Hamiltonian is given by 

\begin{equation}
\hat{\mathrm{H}}_{E} |m,l\rangle = \left [n\omega_{E} + X(m,l) \right ] |m,l\rangle
\end{equation}

\noindent This spreading of the degenerate environment eigenvalues is shown in Figure \ref{fig:ELD} by the gaussians centered on each level.

The composite SE zero-order states $|n\rangle \otimes |m,l\rangle \equiv |n,m,l\rangle$ have off-diagonal elements 

\begin{equation}
\langle n,m,l | \hat{\mathrm{H}}_{SE} | n^\prime, m^\prime, l^\prime \rangle = Y(n,m,l)
\end{equation}

\noindent where $Y(n,m,l)$ is a random variate selected from a gaussian distribution with zero mean and standard deviation   $\sigma=\gamma\omega_{E}$.   (Note that although $\gamma$ is the same, this is a different standard deviation than above.)   To maintain a Hermitian matrix,  we generate random variates only for the upper triangle and map these to the lower triangle.  The resulting Hamiltonian is diagonalized to yield the energy eigenvalues and vectors of SE, which allows for analytic time propagation of the initially selected states.  

\subsubsection{Simulation Basis and ``Microcanonical Shell"}    \label{microcanonicalshell}
  The system basis $\{ | n \rangle \}$  consists of $N_S = 6$  states with $ n = 0 \cdots 5$.  Our environment basis has zero-order quantum numbers $ m = 0 \cdots 7$ with  a degeneracy pattern $6 \times \{1, 2, 4 \cdots 128 \}$ (see discussion above of environment energy pattern and temperature) or $ N_E = \{6 + 12 + 24 + \cdots + 768 \}$ = 1530 states, for a total of $N_{SE} = N_S \times N_E = 9180$ SE basis states.   

\begin{figure}[htb]
\rotatebox{270}{\scalebox{.3}{\includegraphics{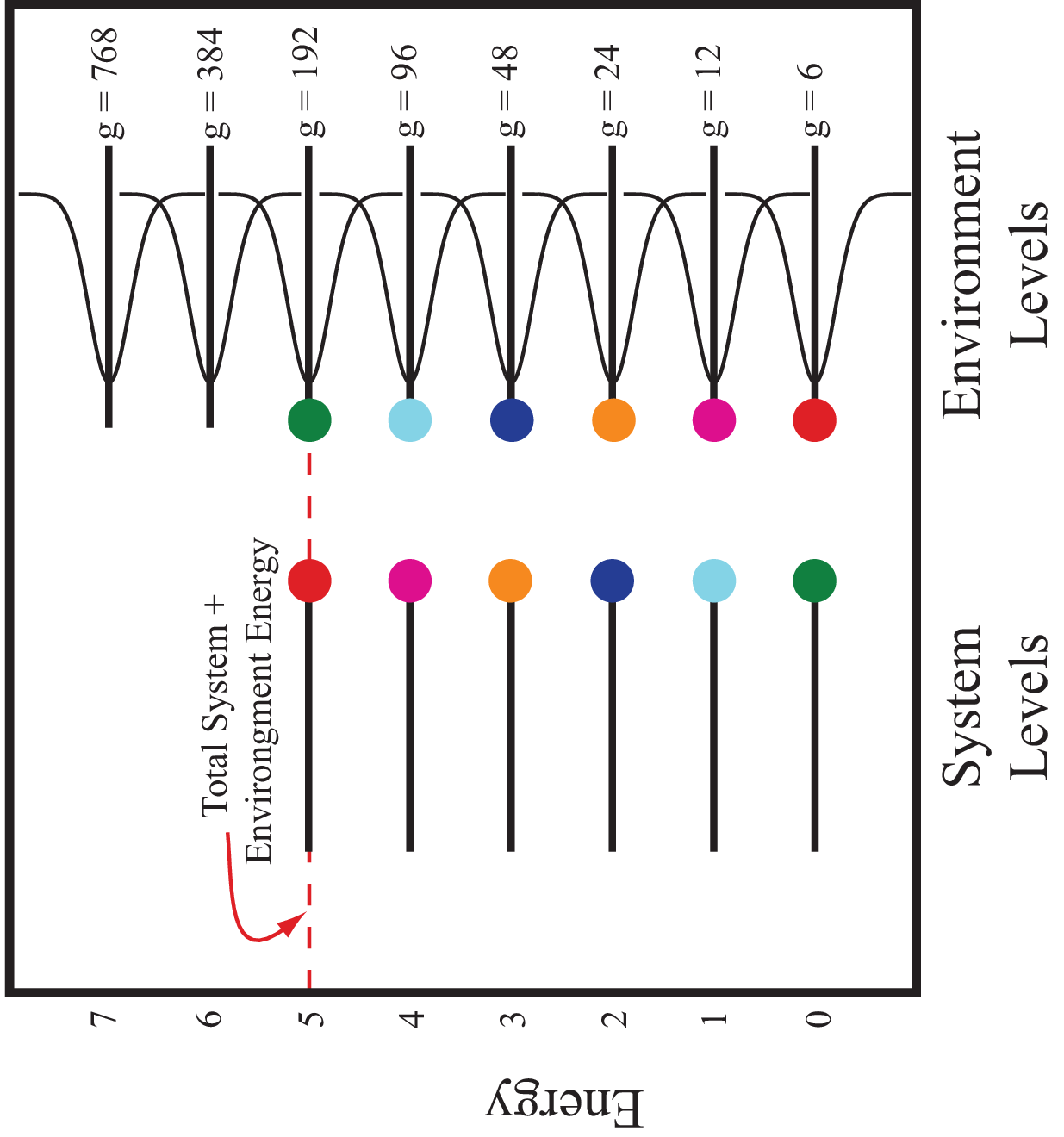}}}
\caption{Energy level diagram and ``microcanonical shell" scheme for the system + environment universe SE, from Ref. \cite{polyadbath}.  There are 6 system eigenlevels and 8 environment energy levels.  The degeneracy scheme for the environment is shown along with a small gaussian spread for the environment levels.   We select initial states of the system and environment making use of the notion of a constant total universe energy $E \sim 5$ (dotted red line) with the initial system state thought of as a fluctuation within the SE universe, enabled by the environment bath.  For a given fluctuated system state (colored dot on system level) the corresponding environment state (matching colored dot on environment level) is chosen in order to conserve the total universe energy. The set of all zero order levels with total energy  $E \sim 5$ constitutes the microcanonical shell.  \label{fig:ELD}}
\end{figure}

We will  work with initial SE states that have a zero-order energy $\sim  5$, thus SE energy quantum numbers with $ n + m = 5$.  Since the S states are singly degenerate with energy = 0 - 5 and the first six E levels have energy $\le 5$, there are a total of 378 SE zero-order states with energy $\sim 5$.  These comprise a ``microcanonical shell" with energy $ E \sim 5$.  This scheme is illustrated in Fig.  \ref{fig:ELD}

\subsubsection{Initial State Selection}

We choose the initial state of SE to be a pure superposition state of the  $|n,m,l\rangle$ basis sates, with equal coefficients for each E basis state.  We do this not because this particular initial state is necessarily more likely than others, but simply as a reflection of our ignorance about the condition of the environment in the initial state.  The equal coefficient assumption may be said to correspond to ``microcanonical" conditions of the environment in the initial fluctuated system state.  We illustrate the balance between the total universe energy with the portion placed in S and E in Figure \ref{fig:ELD}.  For example, if we have a total universe energy of 5 and we wish to place the system initially in the third S energy level ($n=2$, dark blue dot), then in order to maintain the total energy of the universe E must occupy the fourth environment energy level ($m=3$, dark blue dot). All of the quasi-degenerate levels for the fourth E have equal initial probability, which is equivalent to supposing that we are starting our state on a particular energy ``micro-shell'' given our fluctuated S state.   The choice of the total energy in the universe is arbitrary; by picking 5 we are able to excite all levels within S as illustrated by the series of colored circles in Figure \ref{fig:ELD}.  In this work we made use of 6 S eigenlevels and 8 E eigenlevels with the degeneracy scheme in Figure \ref{fig:ELD}.    We note that more E  levels  than S levels are included (8 vs. 6, neglecting degeneracy).  This choice yielded more consistent results for the fitted temperature obtained for each initial state.  One way of thinking about this result is that the ``extra'' E energy levels are needed to converge the calculation.

\subsection{Dynamical Analysis:  Reduced Density Matrix, Entropy, and Free Energy}

We perform analytic time propagation of the initial states, using an expansion in terms of the eigenstates of the SE universe, to obtain the time dependent SE state $| \Psi(t) \rangle $ and from it the reduced density matrix.    We begin with the universe density operator $\rho_{SE}$ and calculate the RDM   $\rho_S=Tr_E \rho_{SE}$ with matrix elements

\begin{equation}
\rho^{n,n^\prime}_{S} = \sum_{m,l} \langle n,m,l|\Psi(t)\rangle\langle\Psi(t)|n^\prime,m,l\rangle
\end{equation}

\noindent  where the summation gives the trace over E.

The calculation of the change in the free energy of the system is based on system properties including the von Neumann entropy, calculated entirely from the RDM of the system $\rho_S$:   

\begin{eqnarray} \Delta F_{sys} &=& \Delta U_{S} - T \Delta S_S^{vN}   \nonumber  \\
 U_S &=& \langle E \rangle_{S}     \label{freeenergy}   \end{eqnarray}

\noindent (Here \textit{sys} refers to the generic system as in Eq.  \ref{greatresult} , and S refers to the model system  as in Eq.  \ref{SEmodel}.)  It is important to note that the free energy change in Eq. \ref{freeenergy} is not a simple sum of changes in system and environment von Neumann entropies, since generally $\Delta S^{vN}_{env} \neq -Q/T = -\Delta U_S/T$ according to the Schmidt decomposition \cite{nielsenchuang}.  We use the temperature $T = 230.41$ K  from the analytic degeneracy formula in Ref. \cite{polyadbath}.  Another possibility would be to use the temperature obtained from fitting the reduced system density to a Boltzmann distribution, a procedure we compared to $T$ in Ref. \cite{polyadbath}.  This gave a fluctuating $T_{fit} $ in reasonably good agreement with $T = 230.41$ K, but using $T_{fit}$ here in the present application gives decidedly inferior results (which we do not report here) to the analytic {$T = 230.41$ K.

\section{Results}      \label{results}

\subsection{Universe Entropy and System Free Energy}

\begin{figure}[h]
\begin{tabular}{c}
\rotatebox{270}{\scalebox{.3}{\includegraphics{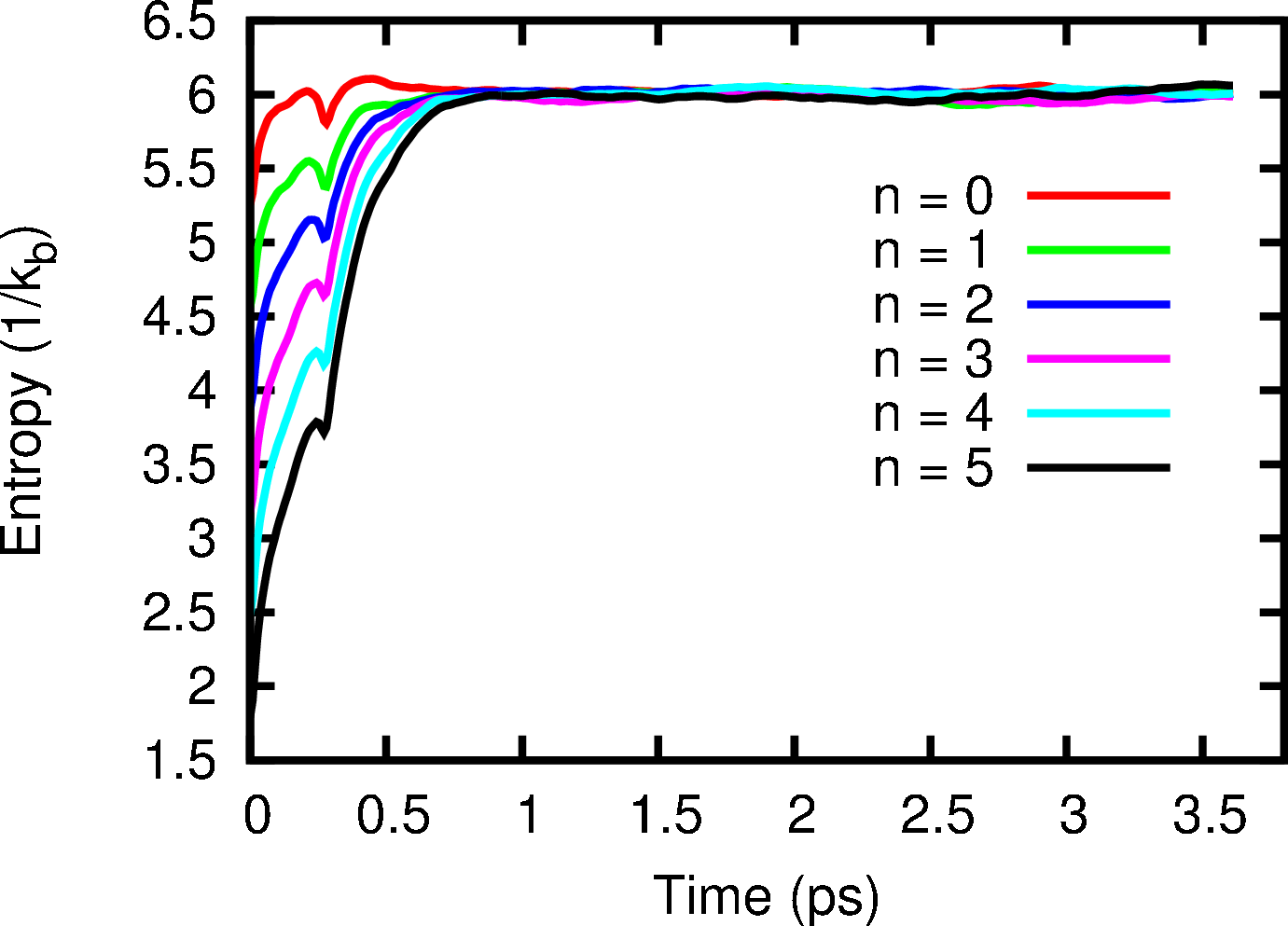}}} \\
\rotatebox{270}{\scalebox{.3}{\includegraphics{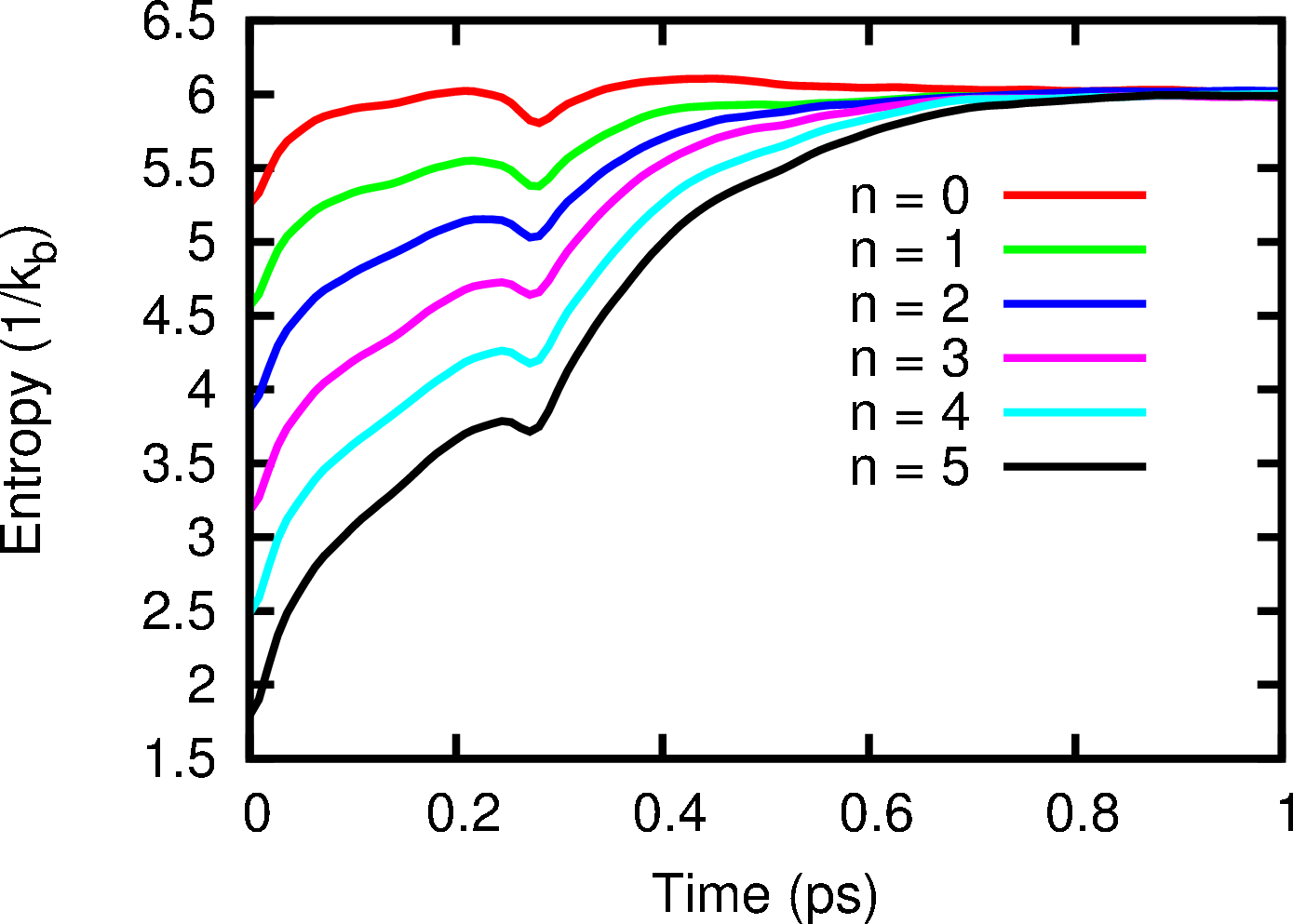}}}
\end{tabular}
\caption{Total entropy of the universe calculated using $-\sum_\alpha p_\alpha \ln (p_\alpha)$ where the $p_\alpha $'s are the populations in the zero order energy basis of the SE universe.  The total universe basis has a size of 9180 (6 system levels times 1530 environment levels).  Note the different time scales in the two panels.    \label{deltasuniv}}
\end{figure}

\begin{figure}[h]
\begin{tabular}{c}
\rotatebox{270}{\scalebox{.3}{\includegraphics{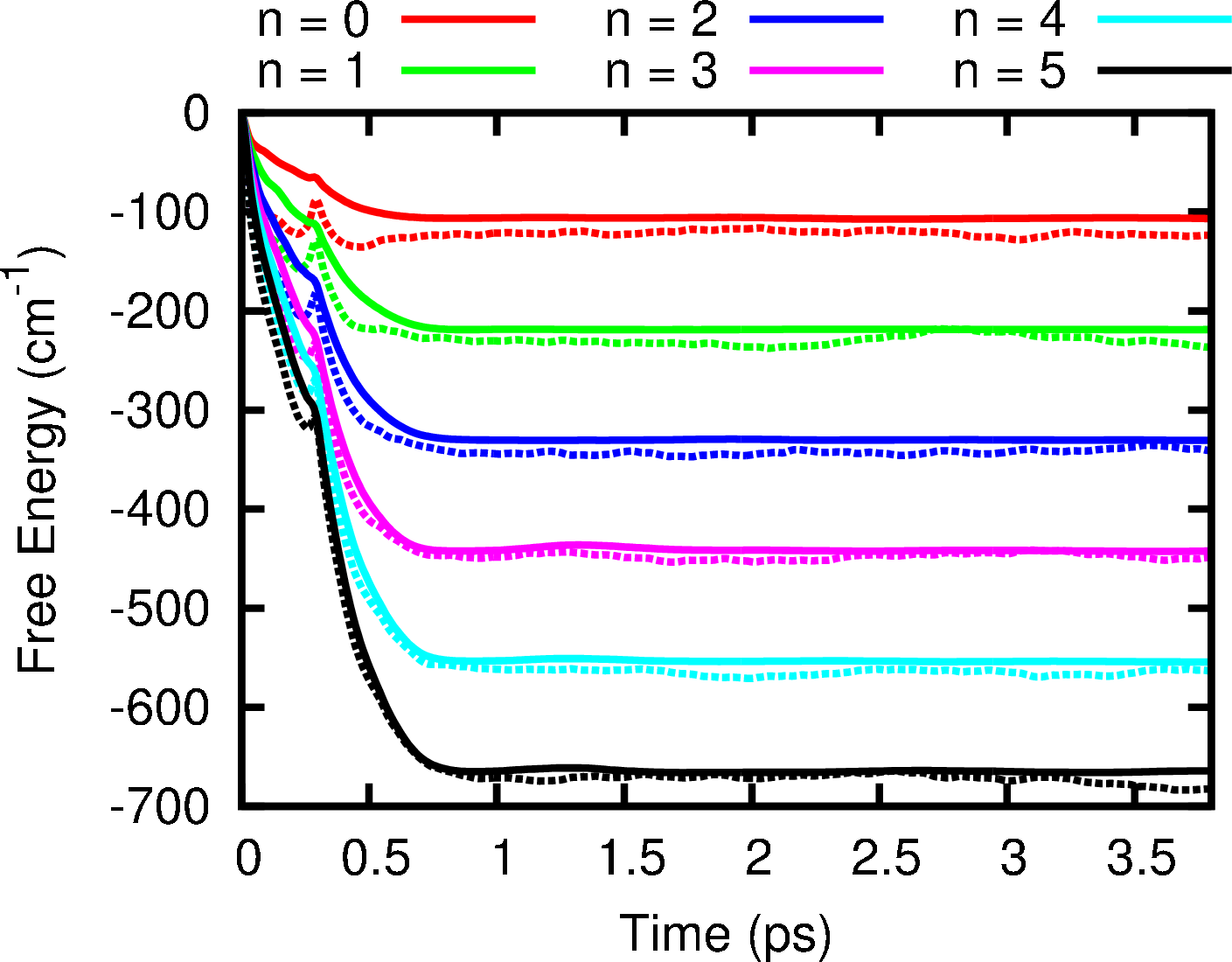}}} \\
\rotatebox{270}{\scalebox{.3}{\includegraphics{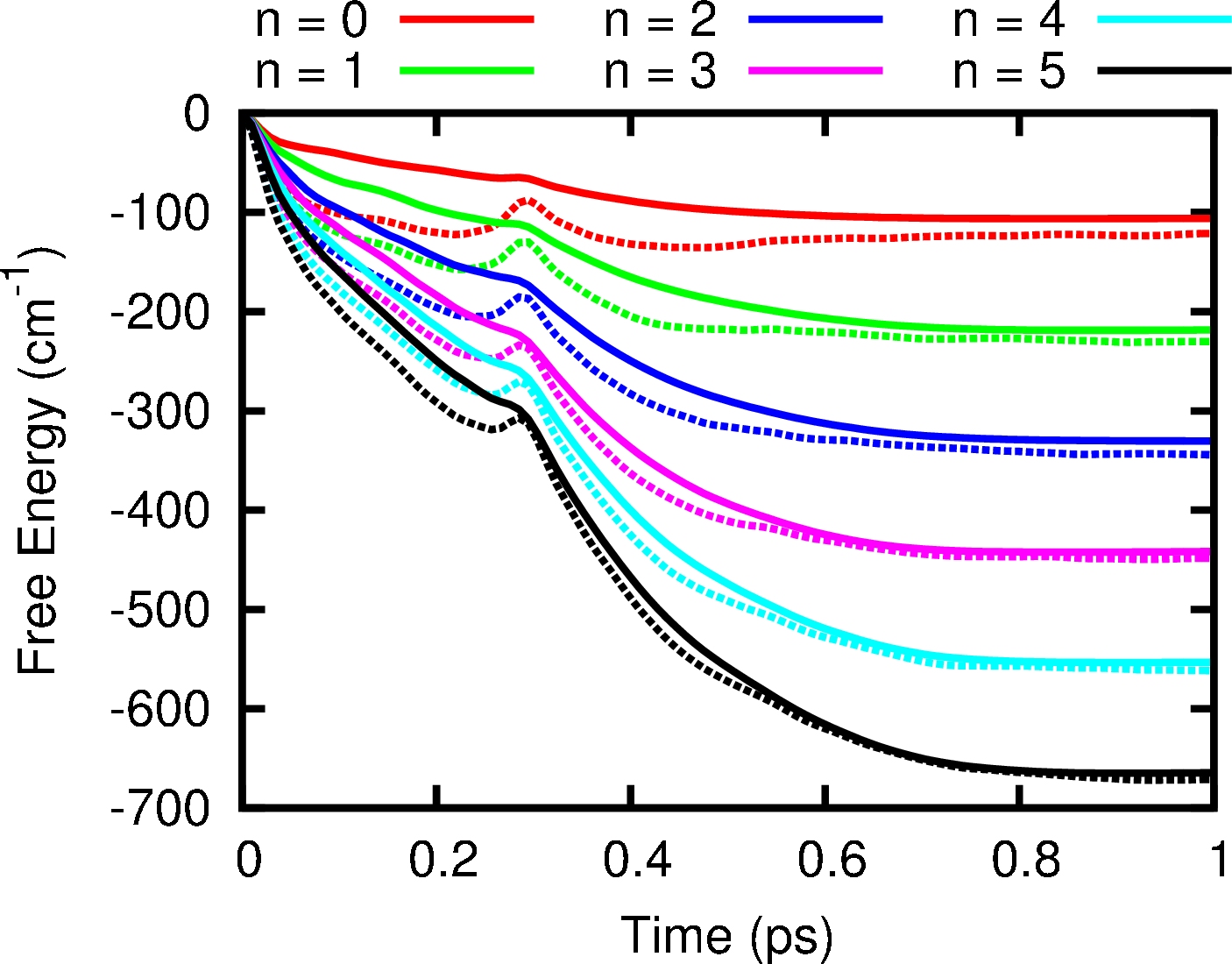}}}
\end{tabular}
\caption{A comparison of two methods of calculating the free energy change.  The solid lines represent $\Delta U_{S}-T\Delta S_S$ with $T=230$ Kelvin.  The dashed lines represent $-T\Delta S^Q_{univ}$ also with $T=230$ Kelvin.    Note the different time scales in the two panels.   \label{free}}  
\end{figure}

Figure \ref{deltasuniv} shows the calculations of   irreversible entropy production $\Delta S^Q_{univ}$ as a function of time for each initial state.  Figure \ref{free} shows comparison of the $\Delta S^Q_{univ}$ with the  $\Delta F_{sys}$  calculations.    In general, they compare well, with some interesting differences along the way evolving in time.   We will briefly consider these anomalies  after first discussing the generally good agreement with Eq. \ref{greatresult} and its connection to the general principles enunciated earlier.

A word is first in order about treating $S^Q_{univ}$ and $F_{sys}$ as time-dependent variables in Figs. \ref{deltasuniv},\ref{free}.   It may be objected that these quantities are not physically defined during the time evolution between the initial state and the final equilibrium.    On the other hand,  it can be maintained that the expression in Eq.  \ref{greatresult} is not just a statement about total changes in thermodynamic quantities in a spontaneous process, but about time derivatives of those quantities in a nonequilibrium system undergoing change.  In the language of Prigogine \cite{kp}, the system expends free energy in irreversible entropy production  $\sigma_i$ which is equivalent to the rate of entropy change of the universe, so Eq. \ref{greatresult} is expressed in differential form as

\begin{equation}    - \frac{1}{T}    \   \dot F_{sys}  =   \sigma_i    =    \dot S^Q_{univ}   > 0    \label{diffgreatresult}  \end{equation} 

\noindent In any case, $S^Q_{univ}$ and $F_{sys}$ are computationally well-defined, as are the values plotted in the figures.

\subsection {     Microcanonical spreading and {state fragmentation}}  \label{fragmentation} 

Earlier we framed our expectations of the computational quantum thermodynamic procedure by appeal to the idea of the microcanonical ensemble of classical statistical mechanics.  Here we examine to what extent a notion of       
  microcanonical behavior is observed in the quantum simulations.   Microcanonical behavior would be observed if through a process of equal ``fragmentation," all of the $p_\alpha$ in Eq. \ref{expansion}  turned out to be equal with $p_\alpha = 1/W$ for the ``microcanonical shell" of $W$ energetically accessible states, and all other $p_\alpha$ were zero  for basis states outside the shell.   However, this seems  a distinctly problematic notion.   As discussed in Section \ref{tmef}, there is no reason whatsoever   to expect a priori that the instantaneous values $p_\alpha = |c_\alpha(t)|^2$ should all be equal to a single value $1/W$.  Our definitions do not involve even implicitly a notion of time averages or {``ergodic"} behavior.  Furthermore, as noted already,  there is no exact notion of a microcanonical shell because the states considered are time-dependent, with no exactly defined energy.  (There would hardly be any point in using the entire SE basis in the simulations if the dynamics were expected to be confined to the basis of the microcanonical shell.)  On the other hand, complete {``fragmentation"} within the entire SE product basis of dimension $N_{SE} = N_S \times N_E$ seems out of the question at less than infinite temperature, and in any case would not correspond to a sensible idea of microcanonical behavior at fixed energy.  

The actual behavior in the simulations tells a story that weaves an interesting pathway amidst these conflicting considerations.  As noted above in Section \ref{microcanonicalshell},  we can define a kind of microcanonical shell of zero order SE states with zero order energies $ E \sim 5$.  We determined that there are 378 such zero order states, out of a total of 9180 states in our full zero order SE basis.  It is  interesting that in the simulations of Fig. \ref{deltasuniv}, we find in all cases a final $S^Q_{univ} \sim 6$.  Exponentiating to get an effective number of states, we find $ e^6 \sim 403$, a pretty good match to a microensemble with 378 states.  (The actual range observed for all initial states is $S^Q_{univ}$ = 5.97 - 6.02.)

\begin{figure}[h]
\begin{tabular}{c}
\rotatebox{270}{\scalebox{.3}{\includegraphics{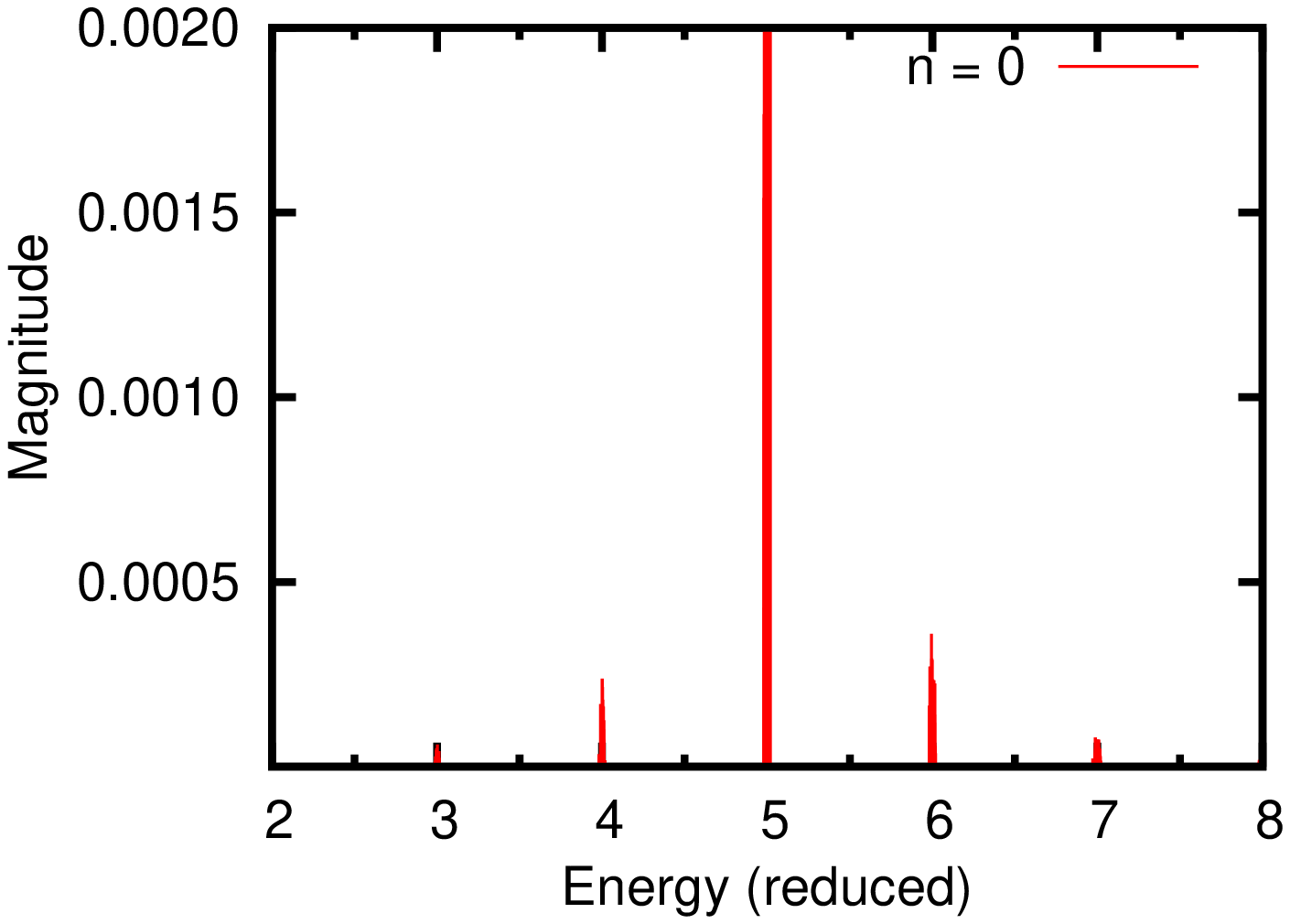}}} \\
\rotatebox{270}{\scalebox{.3}{\includegraphics{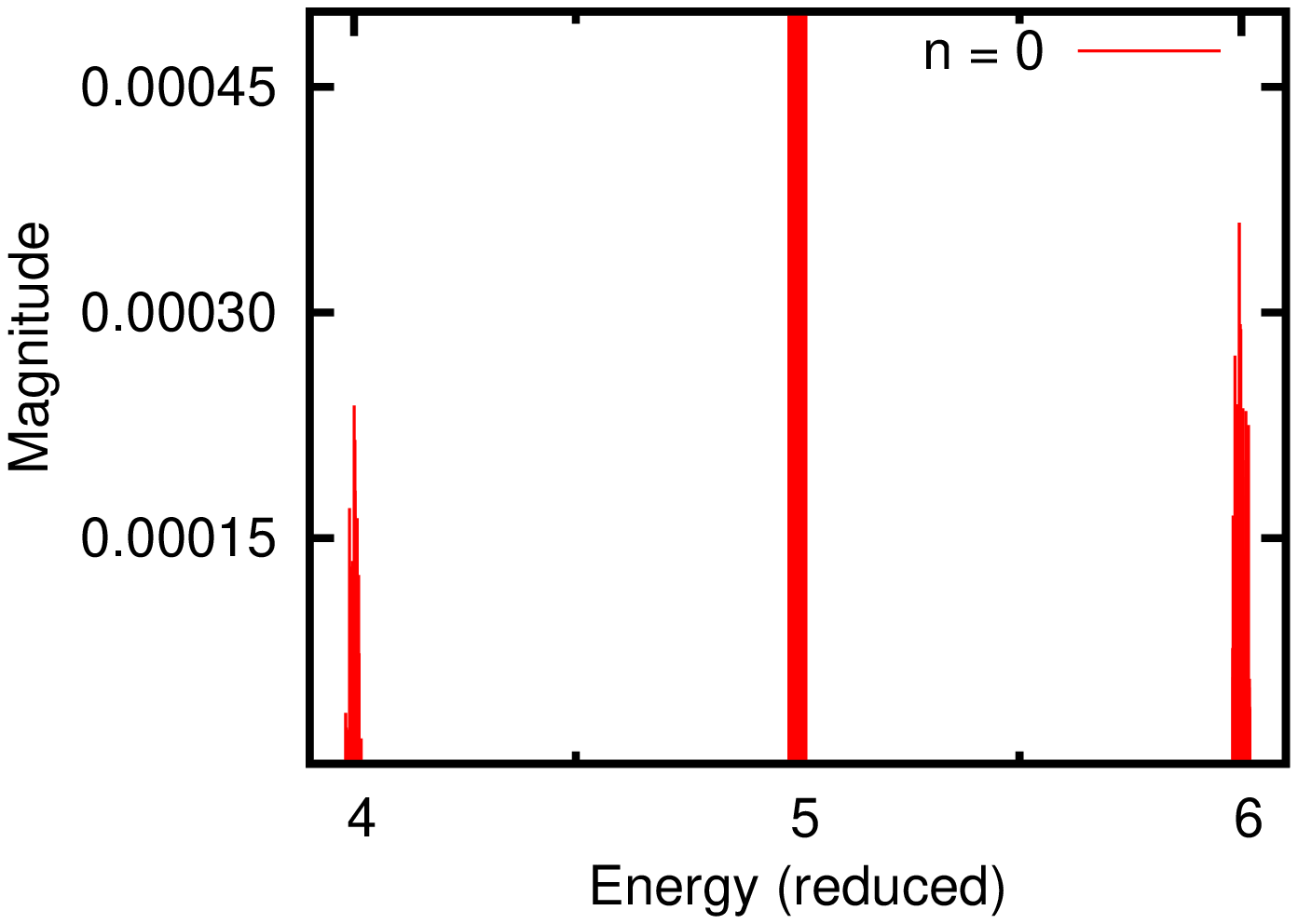}}} \\
\rotatebox{270}{\scalebox{.3}{\includegraphics{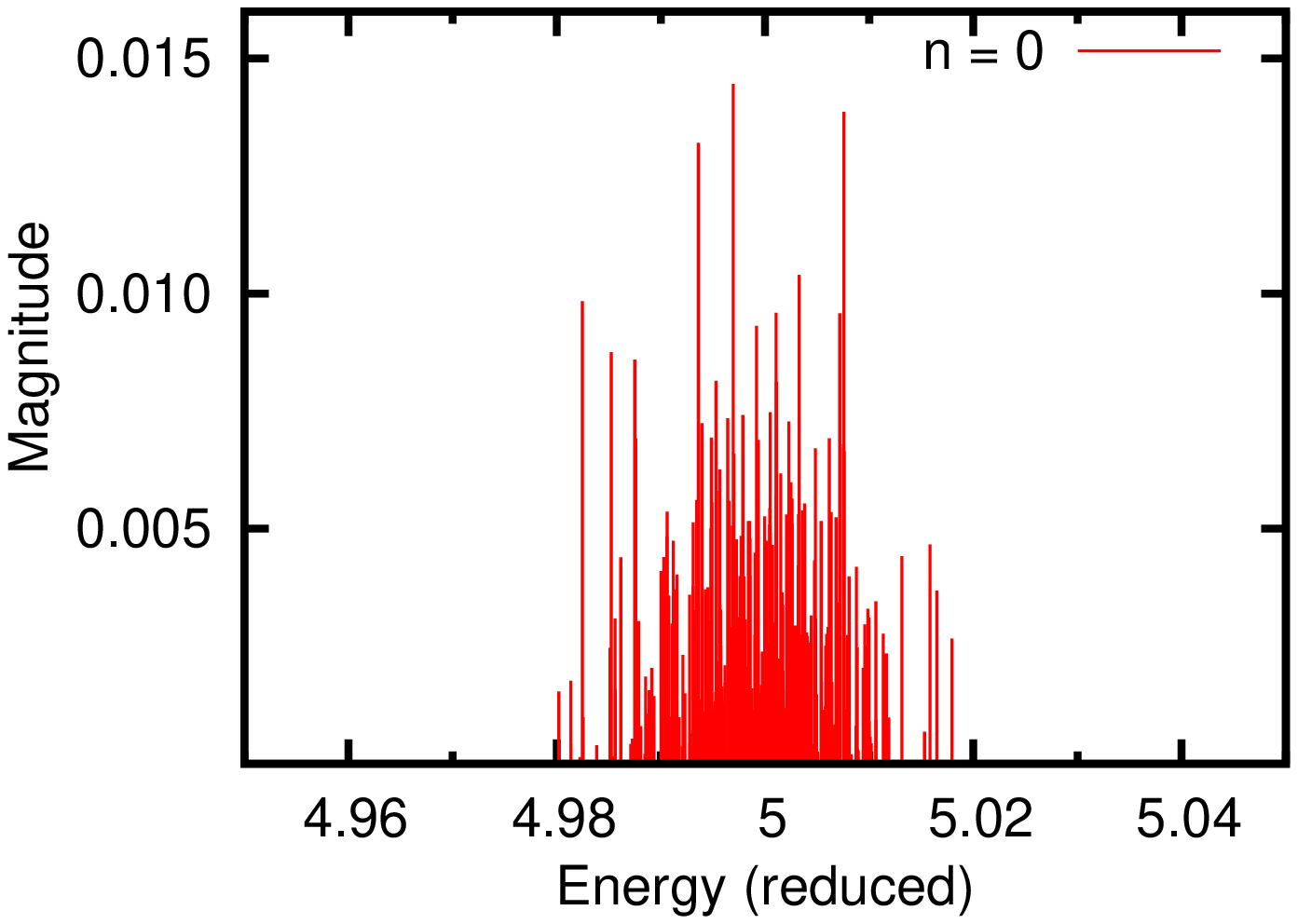}}} \\
\end{tabular}
\caption{A plot of $p_\alpha$ versus the zero order energy for the states with $E \sim 5.0$ for the $n=0$ initial state.  The three panels show various levels of detail discussed in the text.}
\label{sticks}
\end{figure}

 We explore this observation further by examining the $p_\alpha$ of the states of the zero order basis in the expansion (\ref{sticks}) for $ | \Psi (t) \rangle$.  With exactly equal {fragmentation} confined within the E = 5 microcanonical shell, 378 of the zero order states would have $p_\alpha = 1/378$ and all the remaining states of the 9180 member basis would have $p_\alpha = 0$.  What is actually obtained in the simulations is illustrated in Fig. \ref{sticks}.  These are a set of stick diagrams for the entire SE basis for the late-time SE state $| \Psi(t) \rangle$ obtained by time evolving the initial $n = 0$ state in interaction with the environment.  (The diagrams for all the initial states $n = 0 - 5$ are qualitatively very similar).    The stick diagram  shows the magnitudes $p_\alpha$ of the zero order SE states in the final equilibrated  pure SE state $ | \Psi (t) \rangle$  (at a single moment in time -- the data are fluctuating).    In this ``fragmentation" of the final state it can be seen that all the most significant magnitudes are grouped in a narrow band about $E = 5.0$, the notional energy of our microcanonical shell.  (Recall that these zero order states have a spread in zero order energy due to the small spread assumed for the environment states).  Outside of this band of states around $E = 5.0$, all the $p_\alpha$ are  small, but visible in the stick diagram for zero order states of zero order energy $ E \ne 5.0$, especially 4.0 and 6.0.   As we shall see, these small contributions from the other energy microshells are significant.  

The magnitudes of the sticks in the E = 5.0 band show considerable variation.  These are far from being all equal and hence  not very close to ``microcanonical" behavior within the energy shell.  This is shown in fine  detail in the bottom panel of Fig. \ref{sticks}.    It is most interesting that the entropy  calculated from the $p_\alpha$ from the $E = 5.0 $ microcanonical shell according to 

\begin{equation}   S_{shell} =   -\sum_{\alpha, E = 5.0} p_\alpha \ln p_\alpha    \end{equation}

\noindent is not especially close to the observed final $ S^Q_{univ} \sim 6$; instead, it is typically $\sim 5$.  This means the remainder of the entropy $\sim 1$ comes from contributions from the other bands, mainly E = 4.0 and 6.0.  Table \ref{entropymismatch} shows this for all of the initial oscillator energy eigenstates.   The entropy contribution from the E = 5.0 microcanonical shell corresponds to an effective number of states  $e^5 \sim 148$, while the total effective number of states is much larger at  $e^6 \sim 403$.  As noted already, this effective number of states is slightly larger than the 378 of the microcanonical shell.  This can perhaps be understood as an effect of having a time-dependent quantum state.  This is associated with an energy uncertainty, which in turn is associated with an  ``off-shell" effect on the entropy, making it slightly larger than ``microcanonical."  The effect here is consistent with  the small energy uncertainty in our simulations.  If the explanation of the $S^Q_{univ}$ ``entropy excess" offered here is correct, it is an interesting aspect of quantum thermodynamics of time dependent systems that merits further exploration in future work, especially in systems with much stronger time dependence and quantum energy uncertainty.

\begin{table}
\caption{The first column is the total entropy $S^Q_{univ}$ attained for each initial state.  The second column if the partial entropy obtained from the subset of states with the microcanonical energy $E = 5$.  } 

\

\begin{tabular}{ccc}
\hline
$n$ \ \ \ & $S^Q_{univ}$ \ \ \  & $S_{partial}$ \\
\hline\hline
0 & 6.00  &      5.14\\
1 & 6.02  &      5.16\\
2 & 5.97  &      5.08\\
3 & 5.98  &      5.10\\
4 & 5.97  &      5.11\\
5 & 6.00  &      5.14\\
\end{tabular}
\label{entropymismatch}
\end{table}

To summarize:  the $p_\alpha$ for the SE basis states with the microcanonical zero order $E = 5$ are certainly not all equal.  The actual computational sub-entropy from the microcanonical shell does not nearly equal the actual $S^Q_{univ}$.  However, there are small but significant entropy contributions from other microcanonical shells.  Together, all the contributions give a final $ S^Q_{univ} \sim  6$ which is  close to the ``microcanonical" value of $ \ln  378$ for the $E = 5$ shell alone.  This means that we are obtaining an {\it effective} number of microcanonical states that is close to the microcanonical answer from counting the relevant zero order states.  We can say that while equal fragmentation is not obtained within either the microcanonical shell or the complete SE basis, a kind of microcanonical fragmentation occurs within an effective basis, mimicking microcanonical behavior in the numerical value of $S^Q_{univ}$.  In fact, there is a slight excess in entropy which has an explanation in terms of the energy uncertainty associated with time-dependent behavior.

{Our results are very much consistent with arguments \cite{tasaki1998, Reimann2008, Popescu2009, Goldstein2010, Goldstein2015, vNcommentary, vNtrans,Reimann2016} that evolution of a ``typical" pure state leads to an  entangled state congruent with canonical (thermalizing) behavior.  It is not evolution to strict microcanonical entangled state (equal probabilities in all basis states) or time-averaged ergodic behavior that leads to thermalization, but rather the concordance of the probability distribution with a canonical system density.  We saw this in our calculations in the distribution of Fig. 4  of Ref. \cite{polyadbath}. Canonicality also appears in the ``pseudo-microcanonical" count of effective number of states, and even the distributions {\it within} the ``extra" subshells in Fig. \ref{sticks}  outside the energy shell, with slight excess of entropy production.  All of our numerical evidence is consistent with the idea that a single pure quantum state evolves to an entangled universe state that exhibits thermalizing behavior.  }

{\subsection{Anomalies}

In Fig. \ref{deltasuniv} all the initial states evolve to nearly the same final $S^Q_{univ}$ with the desired agreement at long times between the free energy and entropy measures, the key relation of Eq. \ref{greatresult}.   However, along the way there are some anomalies that are worth noting.  These have to do in part with whether the differential version Eq. \ref{diffgreatresult} of the fundamental relation Eq. \ref{greatresult} holds at all times.  Especially in the bottom panel of  Fig. \ref{deltasuniv}, There is a pronounced dip in which the entropy production $ \dot S^Q_{univ} $ becomes negative for a time.  This occurs for all the initial states, at about the same time.  The comparison is also interesting in Fig. \ref{free} of the free energy change $ T \Delta F_{sys}$ with $ - \Delta S^Q_{univ}$, the key relation in Eq. \ref{greatresult}.  Though generally they match reasonably well, especially in the long-time equilibrium limit, for shorter times, there are significant  differences, especially evident in the bottom panel of Fig. \ref{free}.  $S^Q_{univ}$ (which is shown with a minus sign) increases much more rapidly than the free energy  $ - T \Delta F_{sys}$ at short times, before the dip in $ S^Q_{univ}$.  There is no dip in the free energy, only a point of inflection.  It is as if the system is encountering a bottleneck of some kind on the way to equilibrium, reflected as the bump in $ - S^Q_{univ}$.   After surmounting the bottleneck, the free energy declines much more rapidly, before finally flattening out as equilibrium is approached.   These anomalies could be truly interesting, but they might also be mere artifacts, related to the harmonic nature of the energy level pattern of both the system and bath.  These questions merit further exploration in future work on anharmonic systems.

\section{discussion: fundamental issues of quantum thermodynamics and statistical mechanics}

We have found good correspondence in the quantum simulations between the two sides of Eq. \ref{greatresult} for the system free energy and the entropy of the universe, parallel and consistent with fundamental principles of classical thermodynamics.  This has been attained using two very different notions:  an entropy of the universe $S^Q_{univ}$ defined for the pure state of the system-environment universe; and the free energy of the system using the reduced density operator to get the standard von Neumann entropy of the system $S^{vN}$ and the system energy $ \langle E \rangle$.  A way to think about this is that the use of the two different entropies $S^{vN}$ for the system and $S^Q_{univ}$ for the SE universe reflects different informational transformations of S and SE in the spontaneous process.

The system entropy $S^{vN}$ is a measure of  information encoded in the system through entanglement, as represented by the process of tracing over the environment in obtaining the reduced density operator.  
  The universe entropy by contrast is a measure of the information content of the pure state of the universe.  (Here we are following the ideas of Kak \cite{KakEntropy} about information entropy of a pure state, and the conditional information entropy of Stotland et al. \cite{informationentropystotland}.)  The change in the universe entropy $ \Delta S^Q_{univ}$ represents an increase of the complexity of the universe state needed to bring about  a given change in  energy and  information (represented by the von Neumann entropy) of the system.    This information about the universe is not actually known or available without going to the trouble to measure it (as the ``mixing entropy," as discussed in Section \ref{referencebasis}).   In this sense, $\Delta S^Q_{univ}$ also represents an increase in ``unknown information" in a spontaneous process.     The two entropies $S^Q_{univ}$ and $S^{vN}$ can thus be regarded as having  complementary meanings and uses.  

We believe our methods and results are broadly consistent with the program of reconciling the ``ensemblist" and ``individualist" points of view in quantum thermodynamics and statistical mechanics \cite{vNcommentary}. 
In previous work \cite{Gemmerarticle,polyadbath} the compatibility of these approaches was demonstrated in quantum simulations insofar as a thermal distribution is obtained in the process of quantum entanglement of  system and environment for a single pure state.  In this paper we have now demonstrated, in simulations, the compatibility of   the universe pure state property $S^Q_{univ}$ with the  the system property of quantum free energy $F_{sys}$.   

An unexpected aspect of this is the extent to which the simulations depart from a reasonably strict definition of microcanonical behavior within an energy microshell.  It is surprising how much ``off-shell" basis states contribute to the entropy $S^Q_{univ}$, though understandable as an essential aspect of quantum time-dependence.   On the other hand, the entropies obtained agree rather well with the entropy one would predict simply by counting the number of states in the energy microshell, and making  microcanonical ``fragmentation" assumptions.  Effectively, the SE universe mimics microcanonical behavior while playing a bit fast and loose with the rules and definitions.   

In summary, we have found compatibility of the notion of the quantum entropy and free energy of the system with the notion of an entropy of the universe that is defined differently than the von Neumann entropy.  This is in accord with the classical statement of the second law of thermodynamics that the entropy of the universe increases in spontaneous processes.   Whatever the conceptual virtues or shortcomings of our approach to quantum entropy, it gives impressive numerical results.  By this route, it should be possible to use relationships among appropriately defined quantum entropies that mirror and preserve the conventional relationships of standard thermodynamics.  At the same time, this may open up the exploration of novel quantum thermodynamic effects, for example, in exploration of the interplay of quantum thermal effects with the emergence of the classical from the quantum world in quantum entanglement. 

\

\noindent {\bf Acknowledgment}.  This work was supported by the U.S. Department of Energy Basic Energy Sciences program under Contract  DE-FG02-05ER15634.  We would like to thank Steve Hsu for bringing Refs.    \cite{vN,vNtrans,vNcommentary} to our attention, and Benjam\' in Alem\' an, Dietrich Belitz, Jeff Cina,  and John Toner for stimulating discussions.  M.K. especially thanks David Perry for stimulating discussions about the Shannon entropy and quantum thermodynamics.

\bibliography{deltasuniv}

\begin{thebibliography}{40}
\expandafter\ifx\csname natexlab\endcsname\relax\def\natexlab#1{#1}\fi
\expandafter\ifx\csname bibnamefont\endcsname\relax
  \def\bibnamefont#1{#1}\fi
\expandafter\ifx\csname bibfnamefont\endcsname\relax
  \def\bibfnamefont#1{#1}\fi
\expandafter\ifx\csname citenamefont\endcsname\relax
  \def\citenamefont#1{#1}\fi
\expandafter\ifx\csname url\endcsname\relax
  \def\url#1{\texttt{#1}}\fi
\expandafter\ifx\csname urlprefix\endcsname\relax\def\urlprefix{URL }\fi
\providecommand{\bibinfo}[2]{#2}
\providecommand{\eprint}[2][]{\url{#2}}

\bibitem[{\citenamefont{Barnes and Kellman}(2013)}]{polyadbath}
\bibinfo{author}{\bibfnamefont{G.~L.} \bibnamefont{Barnes}} \bibnamefont{and}
  \bibinfo{author}{\bibfnamefont{M.~E.} \bibnamefont{Kellman}},
  \bibinfo{journal}{J. Chem. Phys.} \textbf{\bibinfo{volume}{139}},
  \bibinfo{pages}{21410893} (\bibinfo{year}{2013}).

\bibitem[{\citenamefont{Tasaki}(1998)}]{tasaki1998}
\bibinfo{author}{\bibfnamefont{H.}~\bibnamefont{Tasaki}},
  \bibinfo{journal}{Phys. Rev. Lett.} \textbf{\bibinfo{volume}{80}},
  \bibinfo{pages}{1373} (\bibinfo{year}{1998}).

\bibitem[{\citenamefont{Borowski et~al.}(2003)\citenamefont{Borowski, Gemmer,
  and Mahler}}]{Gemmerarticle}
\bibinfo{author}{\bibfnamefont{P.}~\bibnamefont{Borowski}},
  \bibinfo{author}{\bibfnamefont{J.}~\bibnamefont{Gemmer}}, \bibnamefont{and}
  \bibinfo{author}{\bibfnamefont{G.}~\bibnamefont{Mahler}},
  \bibinfo{journal}{Eur. Phys. J. B} \textbf{\bibinfo{volume}{35}},
  \bibinfo{pages}{255} (\bibinfo{year}{2003}).

\bibitem[{\citenamefont{Gemmer et~al.}(2009)\citenamefont{Gemmer, Michel, and
  Mahler}}]{Gemmer:2009}
\bibinfo{author}{\bibfnamefont{J.}~\bibnamefont{Gemmer}},
  \bibinfo{author}{\bibfnamefont{M.}~\bibnamefont{Michel}}, \bibnamefont{and}
  \bibinfo{author}{\bibfnamefont{G.}~\bibnamefont{Mahler}},
  \emph{\bibinfo{title}{Quantum Thermodynamics: Emergence of Thermodynamic
  Behavior Within Composite Quantum Systems (Second Edition)}}, Lecture Notes
  in Physics (\bibinfo{publisher}{Springer}, \bibinfo{year}{2009}).

\bibitem[{\citenamefont{Popescu et~al.}(2006)\citenamefont{Popescu, Short, and
  Winter}}]{Popescu2006}
\bibinfo{author}{\bibfnamefont{S.}~\bibnamefont{Popescu}},
  \bibinfo{author}{\bibfnamefont{A.~J.} \bibnamefont{Short}}, \bibnamefont{and}
  \bibinfo{author}{\bibfnamefont{A.}~\bibnamefont{Winter}},
  \bibinfo{journal}{Nature Phys.} \textbf{\bibinfo{volume}{2}},
  \bibinfo{pages}{754} (\bibinfo{year}{2006}).

\bibitem[{\citenamefont{Linden et~al.}(2009)\citenamefont{Linden, Popescu,
  Short, and Winter}}]{Popescu2009}
\bibinfo{author}{\bibfnamefont{N.}~\bibnamefont{Linden}},
  \bibinfo{author}{\bibfnamefont{S.}~\bibnamefont{Popescu}},
  \bibinfo{author}{\bibfnamefont{A.~J.} \bibnamefont{Short}}, \bibnamefont{and}
  \bibinfo{author}{\bibfnamefont{A.}~\bibnamefont{Winter}},
  \bibinfo{journal}{Phys. Rev. E} \textbf{\bibinfo{volume}{79}},
  \bibinfo{pages}{061103} (\bibinfo{year}{2009}).

\bibitem[{\citenamefont{Goldstein et~al.}(2006)\citenamefont{Goldstein,
  Lebowitz, Tumulka, and Zangh\`{i}}}]{Goldstein2006}
\bibinfo{author}{\bibfnamefont{S.}~\bibnamefont{Goldstein}},
  \bibinfo{author}{\bibfnamefont{J.~L.} \bibnamefont{Lebowitz}},
  \bibinfo{author}{\bibfnamefont{R.}~\bibnamefont{Tumulka}}, \bibnamefont{and}
  \bibinfo{author}{\bibfnamefont{N.}~\bibnamefont{Zangh\`{i}}},
  \bibinfo{journal}{Phys. Rev. Lett.} \textbf{\bibinfo{volume}{96}},
  \bibinfo{pages}{050403} (\bibinfo{year}{2006}).

\bibitem[{\citenamefont{Goldstein
  et~al.}(2010{\natexlab{a}})\citenamefont{Goldstein, Lebowitz, Tumulka, and
  Zangh\`{i}}}]{vNcommentary}
\bibinfo{author}{\bibfnamefont{S.}~\bibnamefont{Goldstein}},
  \bibinfo{author}{\bibfnamefont{J.~L.} \bibnamefont{Lebowitz}},
  \bibinfo{author}{\bibfnamefont{R.}~\bibnamefont{Tumulka}}, \bibnamefont{and}
  \bibinfo{author}{\bibfnamefont{N.}~\bibnamefont{Zangh\`{i}}},
  \bibinfo{journal}{Eur. Phys. J. H} \textbf{\bibinfo{volume}{35}},
  \bibinfo{pages}{173} (\bibinfo{year}{2010}{\natexlab{a}}).

\bibitem[{\citenamefont{Goldstein
  et~al.}(2010{\natexlab{b}})\citenamefont{Goldstein, Lebowitz, Mastrodonato,
  Tumulka, and Zangh\`{i}}}]{Goldstein2010}
\bibinfo{author}{\bibfnamefont{S.}~\bibnamefont{Goldstein}},
  \bibinfo{author}{\bibfnamefont{J.~L.} \bibnamefont{Lebowitz}},
  \bibinfo{author}{\bibfnamefont{C.}~\bibnamefont{Mastrodonato}},
  \bibinfo{author}{\bibfnamefont{R.}~\bibnamefont{Tumulka}}, \bibnamefont{and}
  \bibinfo{author}{\bibfnamefont{N.}~\bibnamefont{Zangh\`{i}}},
  \bibinfo{journal}{Phys. Rev. E} \textbf{\bibinfo{volume}{81}},
  \bibinfo{pages}{011109} (\bibinfo{year}{2010}{\natexlab{b}}).

\bibitem[{\citenamefont{Goldstein et~al.}(2015)\citenamefont{Goldstein, Hara,
  and Tasaki}}]{Goldstein2015}
\bibinfo{author}{\bibfnamefont{S.}~\bibnamefont{Goldstein}},
  \bibinfo{author}{\bibfnamefont{T.}~\bibnamefont{Hara}}, \bibnamefont{and}
  \bibinfo{author}{\bibfnamefont{H.}~\bibnamefont{Tasaki}},
  \bibinfo{journal}{New J. Phys.} \textbf{\bibinfo{volume}{17}},
  \bibinfo{pages}{045002} (\bibinfo{year}{2015}).

\bibitem[{\citenamefont{von Neumann}(2010)}]{vNtrans}
\bibinfo{author}{\bibfnamefont{J.}~\bibnamefont{von Neumann}},
  \bibinfo{journal}{Eur. Phys. J. H} \textbf{\bibinfo{volume}{35}},
  \bibinfo{pages}{201} (\bibinfo{year}{2010}).

\bibitem[{\citenamefont{Reimann}(2008)}]{Reimann2008}
\bibinfo{author}{\bibfnamefont{P.}~\bibnamefont{Reimann}},
  \bibinfo{journal}{Phys. Rev. Lett.} \textbf{\bibinfo{volume}{101}},
  \bibinfo{pages}{190403} (\bibinfo{year}{2008}).

\bibitem[{\citenamefont{Reimann}(2016)}]{Reimann2016}
\bibinfo{author}{\bibfnamefont{P.}~\bibnamefont{Reimann}},
  \bibinfo{journal}{Nat. Comm.} \textbf{\bibinfo{volume}{7}},
  \bibinfo{pages}{10821} (\bibinfo{year}{2016}).

\bibitem[{\citenamefont{Rigol et~al.}(2008)\citenamefont{Rigol, Dunjko, and
  Olshanni}}]{Rigol}
\bibinfo{author}{\bibfnamefont{M.}~\bibnamefont{Rigol}},
  \bibinfo{author}{\bibfnamefont{V.}~\bibnamefont{Dunjko}}, \bibnamefont{and}
  \bibinfo{author}{\bibfnamefont{M.}~\bibnamefont{Olshanni}},
  \bibinfo{journal}{Nature} \textbf{\bibinfo{volume}{452}},
  \bibinfo{pages}{854} (\bibinfo{year}{2008}).

\bibitem[{\citenamefont{Esposito and Gaspard}(2003)}]{Belgians2}
\bibinfo{author}{\bibfnamefont{M.}~\bibnamefont{Esposito}} \bibnamefont{and}
  \bibinfo{author}{\bibfnamefont{P.}~\bibnamefont{Gaspard}},
  \bibinfo{journal}{Phys. Rev. E} \textbf{\bibinfo{volume}{68}},
  \bibinfo{pages}{066113} (\bibinfo{year}{2003}).

\bibitem[{\citenamefont{Esposito et~al.}(2010)\citenamefont{Esposito,
  Lindenberg, and den Broeck}}]{Belgians}
\bibinfo{author}{\bibfnamefont{M.}~\bibnamefont{Esposito}},
  \bibinfo{author}{\bibfnamefont{K.}~\bibnamefont{Lindenberg}},
  \bibnamefont{and} \bibinfo{author}{\bibfnamefont{C.~V.} \bibnamefont{den
  Broeck}}, \bibinfo{journal}{New J. Phys.} \textbf{\bibinfo{volume}{12}},
  \bibinfo{pages}{013013} (\bibinfo{year}{2010}).

\bibitem[{\citenamefont{Polkovnikov}(2011)}]{PolkovnikovicEntropy}
\bibinfo{author}{\bibfnamefont{A.}~\bibnamefont{Polkovnikov}},
  \bibinfo{journal}{Ann. of Phys.} \textbf{\bibinfo{volume}{326}},
  \bibinfo{pages}{486} (\bibinfo{year}{2011}).

\bibitem[{\citenamefont{Han and Wu}(2015)}]{HanEntropy}
\bibinfo{author}{\bibfnamefont{X.}~\bibnamefont{Han}} \bibnamefont{and}
  \bibinfo{author}{\bibfnamefont{B.}~\bibnamefont{Wu}}, \bibinfo{journal}{Phys.
  Rev. E} \textbf{\bibinfo{volume}{91}}, \bibinfo{pages}{062106}
  (\bibinfo{year}{2015}).

\bibitem[{\citenamefont{Reeb and Wolf}(2014)}]{Reeb}
\bibinfo{author}{\bibfnamefont{D.}~\bibnamefont{Reeb}} \bibnamefont{and}
  \bibinfo{author}{\bibfnamefont{M.~M.} \bibnamefont{Wolf}},
  \bibinfo{journal}{New J. Phys.} \textbf{\bibinfo{volume}{16}},
  \bibinfo{pages}{103011} (\bibinfo{year}{2014}).

\bibitem[{\citenamefont{Almheiri et~al.}(2013)\citenamefont{Almheiri, Marolf,
  Polchinski, and Sully}}]{AMPS}
\bibinfo{author}{\bibfnamefont{A.}~\bibnamefont{Almheiri}},
  \bibinfo{author}{\bibfnamefont{D.}~\bibnamefont{Marolf}},
  \bibinfo{author}{\bibfnamefont{J.}~\bibnamefont{Polchinski}},
  \bibnamefont{and} \bibinfo{author}{\bibfnamefont{J.}~\bibnamefont{Sully}},
  \bibinfo{journal}{J. High Energ. Phys.} \textbf{\bibinfo{volume}{62}}
  (\bibinfo{year}{2013}).

\bibitem[{\citenamefont{Bousso}(2013)}]{Bousso}
\bibinfo{author}{\bibfnamefont{R.}~\bibnamefont{Bousso}},
  \bibinfo{journal}{Phys. Rev. D} \textbf{\bibinfo{volume}{87}},
  \bibinfo{pages}{124023} (\bibinfo{year}{2013}).

\bibitem[{\citenamefont{Hartle and Hawking}(1983)}]{WaveFunctionUniverse}
\bibinfo{author}{\bibfnamefont{J.~B.} \bibnamefont{Hartle}} \bibnamefont{and}
  \bibinfo{author}{\bibfnamefont{S.~W.} \bibnamefont{Hawking}},
  \bibinfo{journal}{Phys. Rev. D} \textbf{\bibinfo{volume}{28}},
  \bibinfo{pages}{2960} (\bibinfo{year}{1983}).

\bibitem[{\citenamefont{Steinhardt and Turok}(2002)}]{steinhardtturok}
\bibinfo{author}{\bibfnamefont{P.~J.} \bibnamefont{Steinhardt}}
  \bibnamefont{and} \bibinfo{author}{\bibfnamefont{N.}~\bibnamefont{Turok}},
  \bibinfo{journal}{Science} \textbf{\bibinfo{volume}{296}},
  \bibinfo{pages}{1436} (\bibinfo{year}{2002}).

\bibitem[{\citenamefont{Rovelli}(2018)}]{Rovelli}
\bibinfo{author}{\bibfnamefont{C.}~\bibnamefont{Rovelli}},
  \emph{\bibinfo{title}{The order of time}} (\bibinfo{publisher}{New York:
  Riverhead Books}, \bibinfo{year}{2018}).

\bibitem[{\citenamefont{von Neumann}(1996)}]{vNbook}
\bibinfo{author}{\bibfnamefont{J.}~\bibnamefont{von Neumann}},
  \emph{\bibinfo{title}{Mathematical Foundations of Quantum Mechanics, Beyer,
  R. T., trans.}}, Princeton Landmarks in Mathematics
  (\bibinfo{publisher}{Princeton University Press}, \bibinfo{year}{1996}).

\bibitem[{\citenamefont{Kondepudi and Prigogine}(1998)}]{kp}
\bibinfo{author}{\bibfnamefont{D.}~\bibnamefont{Kondepudi}} \bibnamefont{and}
  \bibinfo{author}{\bibfnamefont{I.}~\bibnamefont{Prigogine}},
  \emph{\bibinfo{title}{Modern Thermodynamics From Heat Engines to Dissipative
  Structures}} (\bibinfo{publisher}{John Wiley and Sons},
  \bibinfo{year}{1998}).

\bibitem[{\citenamefont{Einstein and Infeld}(1966)}]{EinsteinInfeld}
\bibinfo{author}{\bibfnamefont{A.}~\bibnamefont{Einstein}} \bibnamefont{and}
  \bibinfo{author}{\bibfnamefont{L.}~\bibnamefont{Infeld}},
  \emph{\bibinfo{title}{The evolution of physics from early concepts to
  Relativity and quanta}} (\bibinfo{publisher}{Simon and Schuster},
  \bibinfo{year}{1966}).

\bibitem[{\citenamefont{von Neumann}(1929)}]{vN}
\bibinfo{author}{\bibfnamefont{J.}~\bibnamefont{von Neumann}},
  \bibinfo{journal}{Z Phys.} \textbf{\bibinfo{volume}{57}}, \bibinfo{pages}{30}
  (\bibinfo{year}{1929}).

\bibitem[{\citenamefont{Gibbs}(1902)}]{Gibbs1902}
\bibinfo{author}{\bibfnamefont{J.~W.} \bibnamefont{Gibbs}},
  \emph{\bibinfo{title}{Elementary principles in statistical mechanics}}
  (\bibinfo{publisher}{New York: Charles Scribner's Sons London},
  \bibinfo{year}{1902}).

\bibitem[{\citenamefont{Einstein}(1902)}]{Einstein1902}
\bibinfo{author}{\bibfnamefont{A.}~\bibnamefont{Einstein}},
  \bibinfo{journal}{Annalen der Physik} \textbf{\bibinfo{volume}{314}},
  \bibinfo{pages}{417} (\bibinfo{year}{1902}).

\bibitem[{\citenamefont{Einstein}(1903)}]{Einstein1903}
\bibinfo{author}{\bibfnamefont{A.}~\bibnamefont{Einstein}},
  \bibinfo{journal}{Annalen der Physik} \textbf{\bibinfo{volume}{316}},
  \bibinfo{pages}{170} (\bibinfo{year}{1903}).

\bibitem[{\citenamefont{Einstein}(1904)}]{Einstein1904}
\bibinfo{author}{\bibfnamefont{A.}~\bibnamefont{Einstein}},
  \bibinfo{journal}{Annalen der Physik} \textbf{\bibinfo{volume}{319}},
  \bibinfo{pages}{354} (\bibinfo{year}{1904}).

\bibitem[{\citenamefont{Einstein}(2006)}]{Damour:2006}
\bibinfo{author}{\bibfnamefont{A.}~\bibnamefont{Einstein}},
  \emph{\bibinfo{title}{``On Boltzmann's Principle and Some Immediate
  Consequences Thereof" Translation by B. Duplantier and E. Parks in
  ``Einstein, 1905-2005"}}, Progress in mathematical physics v. 47 Poincare
  Seminar (7th : 2005 : Institut Henri Poincare)" (\bibinfo{publisher}{Boston :
  Birkhäuser Verlag}, \bibinfo{year}{2006}).

\bibitem[{\citenamefont{Kak}(2007)}]{KakEntropy}
\bibinfo{author}{\bibfnamefont{S.}~\bibnamefont{Kak}}, \bibinfo{journal}{Int.
  J. Theo. Phys.} \textbf{\bibinfo{volume}{46}}, \bibinfo{pages}{860}
  (\bibinfo{year}{2007}).

\bibitem[{\citenamefont{Stotland et~al.}(2004)\citenamefont{Stotland,
  Pomeransky, Bachmat, and Cohen}}]{informationentropystotland}
\bibinfo{author}{\bibfnamefont{A.}~\bibnamefont{Stotland}},
  \bibinfo{author}{\bibfnamefont{A.~A.} \bibnamefont{Pomeransky}},
  \bibinfo{author}{\bibfnamefont{E.}~\bibnamefont{Bachmat}}, \bibnamefont{and}
  \bibinfo{author}{\bibfnamefont{D.}~\bibnamefont{Cohen}},
  \bibinfo{journal}{Europhys. Lett.} \textbf{\bibinfo{volume}{67}},
  \bibinfo{pages}{700} (\bibinfo{year}{2004}).

\bibitem[{\citenamefont{Wehrl}(1978)}]{Wehrl}
\bibinfo{author}{\bibfnamefont{A.}~\bibnamefont{Wehrl}}, \bibinfo{journal}{Rev.
  Mod. Phys.} \textbf{\bibinfo{volume}{50}} (\bibinfo{year}{1978}).

\bibitem[{\citenamefont{Nielsen and Chuang}(2000)}]{nielsenchuang}
\bibinfo{author}{\bibfnamefont{M.}~\bibnamefont{Nielsen}} \bibnamefont{and}
  \bibinfo{author}{\bibfnamefont{I.}~\bibnamefont{Chuang}},
  \emph{\bibinfo{title}{Quantum Computation and Quantum Information}}
  (\bibinfo{publisher}{Cambridge}, \bibinfo{year}{2000}).

\bibitem[{\citenamefont{Silvestri et~al.}(2014)\citenamefont{Silvestri, Jacobs,
  Dunjko, and Olshanni}}]{Olshanni}
\bibinfo{author}{\bibfnamefont{L.}~\bibnamefont{Silvestri}},
  \bibinfo{author}{\bibfnamefont{K.}~\bibnamefont{Jacobs}},
  \bibinfo{author}{\bibfnamefont{V.}~\bibnamefont{Dunjko}}, \bibnamefont{and}
  \bibinfo{author}{\bibfnamefont{M.}~\bibnamefont{Olshanni}},
  \bibinfo{journal}{Phys. Rev. E} \textbf{\bibinfo{volume}{89}},
  \bibinfo{pages}{042131} (\bibinfo{year}{2014}).

\bibitem[{\citenamefont{Xiao and Kellman}(1989)}]{Xiao89sphere}
\bibinfo{author}{\bibfnamefont{L.}~\bibnamefont{Xiao}} \bibnamefont{and}
  \bibinfo{author}{\bibfnamefont{M.~E.} \bibnamefont{Kellman}},
  \bibinfo{journal}{J. Chem. Phys.} \textbf{\bibinfo{volume}{90}},
  \bibinfo{pages}{6086} (\bibinfo{year}{1989}).

\bibitem[{\citenamefont{Xiao and Kellman}(1990)}]{Xiao90cat}
\bibinfo{author}{\bibfnamefont{L.}~\bibnamefont{Xiao}} \bibnamefont{and}
  \bibinfo{author}{\bibfnamefont{M.~E.} \bibnamefont{Kellman}},
  \bibinfo{journal}{J. Chem. Phys.} \textbf{\bibinfo{volume}{93}},
  \bibinfo{pages}{5805} (\bibinfo{year}{1990}).

\end{thebibliography}

\end{document}